\documentclass[nofootinbib,aps,prx,longbibliography,twocolumn,superscriptaddress,amsmath,amssymb,verbatim,floatfix]{revtex4-2}
\usepackage[utf8]{inputenc}
\usepackage[english,french]{babel}

\usepackage{bm}
\usepackage{xspace}
\xspaceaddexceptions{]} %
\usepackage{braket} 

\usepackage{siunitx} 
\usepackage[T1]{fontenc}
\DeclareSIUnit[number-unit-product = {}]{\inchQ}{\textquotedbl} 
\DeclareSIUnit[number-unit-product = {\thinspace}]{\inch}{in} 

\usepackage{chemformula} 
\usepackage{graphicx}
\usepackage{lmodern}
\usepackage{epstopdf}
\AtBeginDocument{\usepackage{booktabs}} 
\usepackage[colorlinks=true,citecolor=blue,linkcolor=blue,urlcolor=blue,bookmarks=true]{hyperref}
\usepackage{xcolor}

\newcommand{\mreference}{Ref.~}

\usepackage[nameinlink,capitalize]{cleveref}
\crefname{section}{Sec.}{Secs.}
\crefname{pluralfigure}{Figs.}{Figs.}
\crefname{pluralequation}{Eqs.}{Eqs.}
\creflabelformat{pluralequation}{#2(#1)#3}

\newcommand{\ie}{i.e.\xspace}
\newcommand{\eg}{e.g.\xspace}
\newcommand{\vs}{vs.\xspace}

\newcommand{\via}{via\xspace}
\newcommand{\Via}{Via\xspace}

\newcommand{\apriori}{\textit{a priori}\xspace} 
\newcommand{\abinitio}{\textit{ab initio}\xspace}

\newcommand{\adhoc}{\textit{ad hoc}\xspace} 

\newcommand{\ansatz}{ansatz\xspace}

\newcommand{\nonzero}{non-zero\xspace} 

\newcommand{\nontrivial}{non-trivial\xspace}

\newcommand{\uncalibrated}{uncalibrated\xspace}

\newcommand{\wavefunction}{wavefunction\xspace} 

\newcommand{\planewaveAdj}{plane-wave\xspace}

\newcommand{\zpe}{ZPE\xspace}
\newcommand{\zpeLong}{zero-point energy\xspace}

\newcommand{\zpeLongFirst}{\zpeLong (\zpe)\xspace}

\newcommand{\zpeEq}{\mathrm{ZPE}} 
\newcommand{\zpm}{ZPM\xspace}
\newcommand{\zpmLong}{zero-point motion\xspace}

\newcommand{\zpmLongFirst}{\zpmLong (\zpm)\xspace}

\newcommand{\manybody}{many-body\xspace} 
\newcommand{\Manybody}{Many-body\xspace} 
\newcommand{\ManyBody}{Many-Body\xspace} 

\newcommand{\boltzmann}{Boltzmann\xspace}

\newcommand{\poissonian}{Poissonian\xspace}
\newcommand{\schrodinger}{Schrödinger\xspace}
\newcommand{\zeeman}{Zeeman\xspace}

\newcommand{\bloch}{Bloch\xspace}
\newcommand{\pearson}{Pearson\xspace}
\newcommand{\vdw}{VdW\xspace}
\newcommand{\vdwLong}{van der Waals\xspace}

\newcommand{\vdwLongFirst}{\vdwLong (\vdw)\xspace}

\newcommand{\monkpack}{Monkhorst-Pack\xspace}
\newcommand{\langevin}{Langevin\xspace}

\newcommand{\fourier}{Fourier\xspace}

\newcommand{\lorentzian}{Lorentzian\xspace} 

\newcommand{\dirac}{Dirac\xspace}
\newcommand{\mulliken}{Mulliken\xspace}

\newcommand{\supercell}{supercell\xspace} 

\newcommand{\montecarlo}{Monte Carlo\xspace}
\newcommand{\montecarloAdj}{\montecarlo} 

\newcommand{\dft}{DFT\xspace}
\newcommand{\dftLong}{density functional theory\xspace}

\newcommand{\dftLongFirst}{\dftLong (\dft)\xspace}

\newcommand{\dftd}{\dft-D\xspace}
\newcommand{\xcLong}{exchange--correlation\xspace}

\newcommand{\xcFuncLong}{\xcLong functional\xspace}

\newcommand{\lda}{LDA\xspace}
\newcommand{\ldaLong}{local density approximation\xspace}

\newcommand{\ldaLongFirst}{\ldaLong (\lda)\xspace}

\newcommand{\pbe}{PBE\xspace}
\newcommand{\ts}{TS\xspace}
\newcommand{\tsLong}{Tkatchenko--Scheffler\xspace}
\newcommand{\tsLongFirst}{\tsLong (\ts)\xspace}

\newcommand{\scf}{SCF\xspace}
\newcommand{\scfLong}{self-consistent field\xspace}

\newcommand{\scfLongFirst}{\scfLong (\scf)\xspace}

\newcommand{\castep}{\mbox{CASTEP}\xspace} 
\newcommand{\mufinder}{\mbox{MuFinder}\xspace} 
\newcommand{\calcalc}{\mbox{CalcALC}\xspace} 

\newcommand{\md}{MD\xspace}
\newcommand{\mdLong}{molecular dynamics\xspace}

\newcommand{\pimd}{PIMD\xspace}
\newcommand{\pimdLong}{path-integral \mdLong}
\newcommand{\PimdLong}{Path-integral \mdLong}
\newcommand{\pimdLongFirst}{\pimdLong (\pimd)\xspace}
\newcommand{\PimdLongFirst}{\PimdLong (\pimd)\xspace}

\newcommand{\dftpimd}{{\dft}+{\pimd}\xspace} 

\newcommand{\miwLongAdj}{many-interacting-worlds\xspace}

\newcommand{\dmat}{DM\xspace}
\newcommand{\dmatLong}{dynamical matrix\xspace}

\newcommand{\dmatLongFirst}{\dmatLong (\dmat)\xspace}

\newcommand{\qlcr}{QLCR\xspace}
\newcommand{\qlcrLong}{quadrupolar level-crossing resonance\xspace}

\newcommand{\qlcrLongFirst}{\qlcrLong (\qlcr)\xspace}

\newcommand{\muon}{$\mu$\xspace}
\newcommand{\muonPlus}{$\mu^+$\xspace}
\newcommand{\muoniumEq}{\mathrm{Mu}} 

\newcommand{\musr}{{\muon}SR\xspace} 
\newcommand{\musrLongRelax}{muon spin relaxation\xspace}

\newcommand{\musrLongRelaxFirst}{\musrLongRelax (\musr)\xspace}

\newcommand{\efg}{EFG\xspace}
\newcommand{\efgs}{EFGs\xspace}
\newcommand{\efgLong}{electric field gradient\xspace}

\newcommand{\efgsLong}{electric field gradients\xspace}

\newcommand{\efgLongFirst}{\efgLong (\efg)\xspace}

\newcommand{\efgsLongFirst}{\efgsLong (\efg)\xspace}

\newcommand{\nqcc}{NQCC\xspace}
\newcommand{\nqccLong}{nuclear quadrupolar coupling constant\xspace}

\newcommand{\nqccEq}{C_\mathrm{Q}} 
\newcommand{\nqccStatic}{static \nqcc}

\newcommand{\nqccStaticLong}{static \nqccLong}
\newcommand{\NqccStaticLong}{Static \nqccLong}
\newcommand{\nqccStaticLongFirst}{\nqccStaticLong (\nqcc)\xspace} 
\newcommand{\NqccStaticLongFirst}{\NqccStaticLong (\nqcc)\xspace} 
\newcommand{\nqccStaticEq}{\nqccEq^0} 
\newcommand{\nqccStaticEqDFT}{\nqccEq^{0,\mathrm{\dft}}} 
\newcommand{\nqccStaticEqLongInline}{e_0^2 q_0 Q/h} 
\newcommand{\nqccStaticEqLongInlineFirst}{\nqccStaticEq = \nqccStaticEqLongInline} 
\newcommand{\nqccZeroTempOrderParamEq}{f(0)} 
\newcommand{\nqccCaliFactEq}{f_\mathrm{cal}} 

\newcommand{\muonnuclear}{muon--nuclear\xspace}

\newcommand{\None}{\ch{N}\xspace}
\newcommand{\Ntwo}{\ch{N2}\xspace}
\newcommand{\aNtwo}{$\alpha$--\Ntwo}

\newcommand{\bNtwo}{$\beta$--\Ntwo}

\newcommand{\NtwomuNtwo}{\ch{[N2-{\muon}-N2]+}\xspace} 
\newcommand{\Nisotope}{\ch{^{14}N}\xspace}

\newcommand{\PTwoOneThree}{$P2_{1}3$\xspace}
\newcommand{\PTwoOneThreeFirst}{\PTwoOneThree (No. 198)\xspace}
\newcommand{\PaThree}{$Pa\bar{3}$\xspace}
\newcommand{\PaThreeFirst}{\PaThree (No. 205)\xspace}

\newcommand{\fccLong}{face-centered cubic\xspace}

\newcommand{\chiSqEq}{\chi^2} 
\newcommand{\chiSq}{$\chiSqEq$\xspace}
\newcommand{\reducedChiSqEq}{\chi^2/\mathrm{DOF}} 

\newcommand{\HThreeS}{\ch{H3S}\xspace}
\newcommand{\LaHTen}{\ch{LaH10}\xspace}
\newcommand{\CSHEight}{\ch{CSH8}\xspace}
\newcommand{\LuHThreeN}{\ch{LuH_{3-$\delta$}N_{$\epsilon$}}\xspace}
\newcommand{\hydrogenStorageCHCompound}{\ch{(CH4)3(H2)25}\xspace}

\newcommand{\kB}{k_{\mathrm{B}}} 
\newcommand{\kBLong}{\boltzmann constant\xspace}

\newcommand{\nCubed}[1]{$#1 \times #1 \times #1$\xspace}

\newcommand{\twoCubed}{\nCubed{2}}

\newcommand{\hamiltonian}{Hamiltonian\xspace} 

\newcommand{\hamiltonianEq}{H} 

\newcommand{\approxproptoinner}[2]{%
  \mathrel{%
    \setbox0=\hbox{$#1\sim$}%
    \setbox2=\hbox{%
      \rlap{\hbox{$#1\propto$}}%
      \lower1.1\ht0\box0%
    }%
    \raise0.25\ht2\box2%
  }%
}

\newcommand{\inlinefrac}[2]{#1/#2} 

\newcommand*\dd{\mathop{}\!\mathrm{d}} 

\newcommand{\abs}[1]{\left| #1 \right|} 

\newcommand{\myvec}[1]{{\boldsymbol{\mathbf{#1}}}} 
\newcommand{\mytensor}[1]{#1} 
\newcommand{\normvec}[1]{\myvec{\hat{#1}}} 

\newcommand{\tran}{^{\mathrm{T}}} 

\newcommand{\id}{\mathrm{id}} 
\newcommand{\idtensor}{\mytensor{\id}} 

\DeclareMathOperator{\tr}{tr}

\definecolor{darkgreen}{RGB}{0,128,0}
\definecolor{darkblue}{RGB}{0,0,128}
\definecolor{nblue2}{RGB}{24,118,178}
\definecolor{nyellow}{RGB}{205,116,0} 

\usepackage[normalem]{ulem} 

\usepackage{orcidlink} 

\setcitestyle{super}

\begin{document}

\makeatletter 
\renewcommand\@biblabel[1]{#1.} 
\makeatother

\selectlanguage{english}

\preprint{APS/123-QED}

\title{\ManyBody Quantum Muon Effects and Quadrupolar Coupling in Solids}

\author{Matja\v{z} Gomil\v{s}ek\,\orcidlink{0000-0002-9152-8905}}
\email[Corresponding author. Email: ]{matjaz.gomilsek@ijs.si} 
\affiliation{Jo\v{z}ef Stefan Institute, Jamova c.~39, SI-1000 Ljubljana, Slovenia}
\affiliation{Faculty of Mathematics and Physics, University of Ljubljana, Jadranska u. 19, SI-1000 Ljubljana, Slovenia}
\affiliation{Department of Physics, Durham University, South Road, Durham DH1 3LE, United Kingdom}
\author{Francis L. Pratt\,\orcidlink{0000-0002-5919-3885}}
\affiliation{ISIS Muon Group, Science and Technology Facilities Council (STFC), Didcot OX11 0QX, United Kingdom}
\author{Stephen P. Cottrell\,\orcidlink{0000-0002-8021-6607}}
\affiliation{ISIS Muon Group, Science and Technology Facilities Council (STFC), Didcot OX11 0QX, United Kingdom}
\author{Stewart J. Clark\,\orcidlink{0000-0003-4792-7738}}
\affiliation{Department of Physics, Durham University, South Road, Durham DH1 3LE, United Kingdom}
\author{Tom Lancaster\,\orcidlink{0000-0002-6714-4215}}
\affiliation{Department of Physics, Durham University, South Road, Durham DH1 3LE, United Kingdom}

\date{\today}

\begin{abstract}
Strong quantum \zpmLongFirst of light nuclei and other particles is a crucial aspect of many state-of-the-art quantum materials. 
However, it has only recently begun to be explored from an \abinitio perspective, through several competing approximations. 
Here we develop a unified description of muon and light nucleus \zpm and establish the regimes of anharmonicity and positional quantum entanglement where different approximation schemes apply. 
\Via \dftLong and \pimdLong simulations we demonstrate that in solid nitrogen, \aNtwo, muon \zpm is both strongly anharmonic and \manybody in character, with the muon forming an extended electric-dipole polaron around a central, quantum-entangled \NtwomuNtwo complex. 
By combining this quantitative description of quantum muon \zpm with precision muon \qlcrLong experiments, we independently determine the static \Nisotope \nqccLong of pristine \aNtwo to be $\SI{-5.36(2)}{\mega\hertz}$, a significant improvement in accuracy over the previously-accepted value of $\SI{-5.39(5)}{\mega\hertz}$, and a validation of our unified description of light-particle \zpm. 
\end{abstract}

\maketitle

\renewcommand{\figurename}{\bf Fig.} 
\renewcommand{\tablename}{\bf Table} 

\renewcommand\thetable{\arabic{table}}


\section*{Introduction}
\vspace{-1.25em} %
Quantum \zpmLongFirst of nuclei plays a pivotal role in the structure and dynamics of many important classes of materials, especially those containing light atoms such as hydrogen or lithium~\cite{herrero2014path,ceriotti2016nuclear,markland2018nuclear}. 
Prominent examples of this include recent record high-$T_\mathrm{c}$ hydride superconductors~\cite{drozdov2015conventional,drozdov2019superconductivity,dasenbrock2023evidence,ming2023absence,snider2020room,snider2022retraction,eremets2022high}, record-density hydrogen storage materials~\cite{ranieri2022formation}, metallic and solvated Li and F~\cite{duignan2017real,ackland2017quantum}, as well as many hydrogen-bonded materials~\cite{ceriotti2016nuclear}, \eg, water ice. 
Outside nuclear quantum effects, \zpm should be even more pronounced for implanted muons \muonPlus, which act as sensitive probes in \musrLongRelaxFirst experiments~\cite{blundell1999spin,blundell2021muon,yaouanc2011muon}, since a muon has just 1/9 of the proton mass. 
We expect a large muon \zpeLong $\zpeEq \propto m_\mu^{-1/2} \sim \SI{0.7}{\electronvolt}$~\cite{blundell2021muon,moller2013playing,bonfa2016toward,huddart2020muon,manas2021quantum,prando2013common,onuorah2019quantum}, corresponding \zpm delocalization ${\Delta x} \propto m_\mu^{-1/4} \sim \SI{0.2}{\angstrom}$ around isolated muon stopping sites, the merging of candidate muon sites separated by low energy barriers, quantum tunneling, diffusion, and even muon \bloch waves~\cite{storchak1998quantum,herrero2007diffusion}. 
Beyond these, \manybody quantum effects like positional entanglement between muons and nuclei are also expected. 
Muon \zpm challenges the approach of predicting muon stopping sites and the lattice distortions around them using \abinitio methods, often based on \dftLongFirst~\cite{moller2013playing,bonfa2016toward,huddart2020muon,huddart2022mufinder}, that treats the muon and nuclei as classical particles. 
Several schemes for approximating muon \zpm have been developed~\cite{bonfa2016toward,huddart2020muon}, mainly divisible into:
(i)~adiabatic methods, based on a single-particle description~\cite{soudackov1999removal,porter1999muonium,bonfa2015efficient,onuorah2019quantum,monacelli2021stochastic}, and
(ii)~harmonic approximations~\cite{boxwell1993ab,moller2013quantum,manas2021quantum}. 
Studies of quantum muons using computationally more demanding, but numerically-exact, \pimdLongFirst have remained rather sparse~\cite{herrero2007diffusion,yamada2014accurate,herrero2014path}, despite its popularity in describing light-nuclei systems~\cite{herrero2014path,ceriotti2016nuclear,markland2018nuclear}. 

\begin{figure}[!t]
\centering
\includegraphics[width=1\columnwidth]{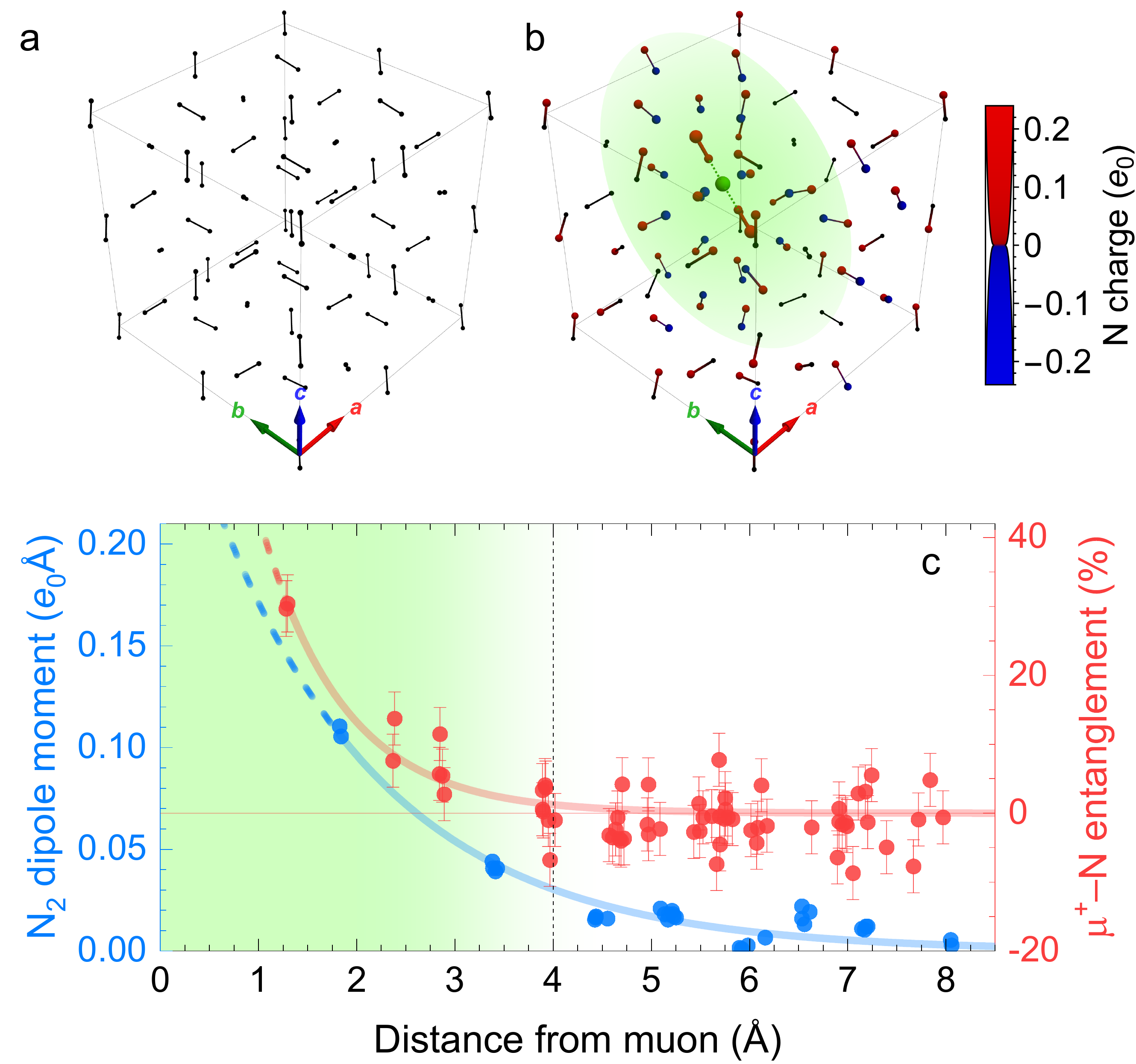}
\caption{%
{\bf Muon site and \muonnuclear entanglement.} 
{\bf a} Pristine \aNtwo crystal structure in a \twoCubed \supercell. 
See also Supplementary Movie~1. 
{\bf b} Classical muon \muonPlus (green) in \aNtwo induces an electric-dipole polaron (shading) around a linear \NtwomuNtwo complex. 
Sphere size/color gives the induced 
\mulliken charge of nitrogen ions. 
See also Supplementary Movie~2. 
{\bf c} Induced \Ntwo \mulliken electric dipole moments (blue) and degree of \muonnuclear positional entanglement $c$ [\cref{eq_corr}] from \pimdLong (\pimd; red). 
Solid lines are exponential fits, dashed line/shading indicates the polaron size. 
$c$ error bars represent a systematic uncertainty of one standard deviation (see Methods). 
}
\label{fig_site_polaron_ent}
\end{figure}

Here we develop a unified description of muon and light nuclei \zpm and the regimes in which particular approximations apply, \via a \dftpimd study of muon \zpm in solid nitrogen, \aNtwo (\cref{fig_site_polaron_ent}a and Supplementary Fig.~1a). 
We characterize the regimes by the degree of (i)~anharmonicity and (ii)~\muonnuclear positional entanglement, the latter quantified with easy-to-calculate entanglement witnesses~\cite{plenio2007introduction,horodecki2009quantum,horodecki2021quantum}. 
We find that in \aNtwo anharmonic \manybody quantum effects dominate, as anticipated from previous experimental~\cite{storchak1998quantum} and theoretical work~\cite{claxton1995muon}, and discover that an extended electric-dipole polaron of polarized \Ntwo molecules forms around a central \NtwomuNtwo complex (\cref{fig_site_polaron_ent}b). 
By applying these insights to our precision muon \qlcrLongFirst~\cite{abragam1984spectrometrie,storchak1992muon,lord2014time} experiments, we derive an independent estimate of the \nqccStaticLongFirst of \Nisotope with significantly improved accuracy 
compared to the literature value~\cite{scott1976solid}. 

\vspace{-1.5em} %
\section*{Results}
\vspace{-1.25em} %
%
\subsection*{Classical muon}
\vspace{-1.25em} %
Within the classical, point-particle description of muons and nuclei using \dft, we find a single stable muon site at almost exactly the $(0, \inlinefrac{1}{4}, \inlinefrac{1}{4})$ position (\cref{fig_site_polaron_ent}b and Supplementary Fig.~1b), which lies between two molecules of pristine \aNtwo but is not symmetric under its \PaThree space group~\cite{erba2011post,rumble2021crc}. 
In line with \abinitio simulations of muons~\cite{claxton1995muon} and protons~\cite{botschwina2001theoretical,terrill2010ab,yu2015structure,liao2017infrared,hooper2019assignment} in \Ntwo clusters, we find that in crystalline \aNtwo a muon forms an almost-linear, almost-centrosymmetric \NtwomuNtwo complex oriented along the $[0, 1, 1]$ direction. 
Within this complex the muon's bare positive charge $+e_0$ is screened to just $+0.46 e_0$ by electrons covalently shared with the two nearest \Ntwo molecules, leaving them with an electron density deficit and thus a net positive charge of $+0.27 e_0$ each ($+0.04 e_0$ on the two nearest \None atoms and $+0.23 e_0$ on the two further-away ones; \cref{fig_site_polaron_ent}b and Supplementary Fig.~2). 
Moreover, we find that an unusual, extended electric-dipole polaron forms around this complex, where its positive charge further induces electric dipole moments (\cref{fig_site_polaron_ent}c) on other, net neutral nearby \Ntwo molecules (Supplementary Fig.~2) and causes them to reorient to point towards the complex. 
Up to $\SI{\sim 4}{\angstrom}$ from the muon, dipolar electrostatic interactions of polarized \Ntwo molecules with the \NtwomuNtwo complex thus overwhelm the weak electric quadrupole and \vdwLongFirst \Ntwo--\Ntwo interactions of pristine \aNtwo~\cite{scott1976solid,storchak1999muonium}. 

\vspace{-1.5em} %
\subsection*{Single-particle quantum approximations}
\vspace{-1.25em} %
\begin{figure}[!t]
\centering
\includegraphics[width=1\columnwidth]{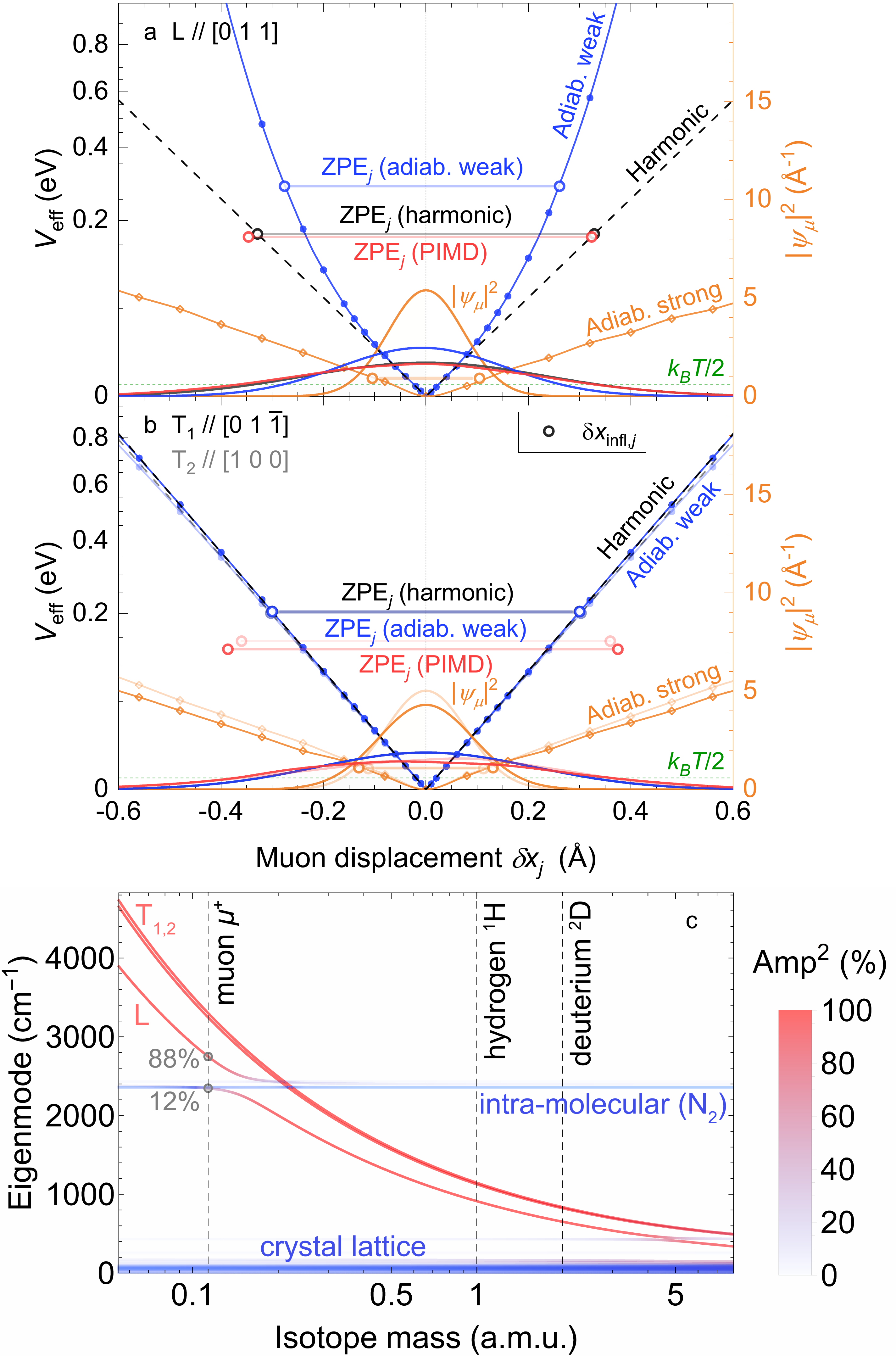}
\caption{%
{\bf Effective muon potential and vibrational spectra.} 
{\bf a}, {\bf b} Effective muon potential in a square-root--linear scale (harmonic potentials thus appear as straight lines) along the {\bf a} $\mathrm{L}$ and {\bf b} $\mathrm{T}_{1,2}$ directions for various approximations (left axis) with the corresponding 
muon probability density function (right axis), directional \zpeLongFirst contributions $\zpeEq_j$, and muon \wavefunction inflection points ${\delta x}_{\mathrm{infl},j}$, where $\partial^2 \psi_\mu / \partial ({\delta x}_j)^2 = 0$. 
Under the harmonic approximation, the muon \wavefunction width is ${\Delta x}_j = {\delta x}_{\mathrm{infl},j}/\sqrt{2}$ (see also \cref{table_zpm_parameters}). 
For \pimdLongFirst calculations (at $T = \SI{20}{\kelvin}$), $\zpeEq_j$ is estimated as the harmonic $\zpeEq_j$ yielding the same ${\Delta x}_j$ if $m_\mathrm{eff} = m_\mu$ (\cref{table_zpm_parameters}). 
{\bf c} $\Gamma$-point phonon spectra of a \supercell of muonated \aNtwo \vs muon isotope mass $m_\mu$ [plot generated from a single \dftLongFirst calculation; see Methods]. 
Shading corresponds to the muon amplitude squared in individual normal modes. 
}
\label{fig_potential}
\end{figure}

To incorporate quantum effects we first employ single-particle, adiabatic approximations of \muonnuclear \zpm. 
Though elaborate schemes have been proposed~\cite{onuorah2019quantum,monacelli2021stochastic}, the simplest are weakly- and strongly-bound muon approximations~\cite{soudackov1999removal,porter1999muonium,bonfa2015efficient}, corresponding to zero or maximal \muonnuclear entanglement, respectively. 
In both schemes, an effective single-particle muon potential $V_\mathrm{eff}(\myvec{\delta x})$ is constructed from total \dft energy under muon displacements $\myvec{\delta x}$ from its classical site, while: 
(i)~keeping the nuclei fixed at the positions corresponding to the unperturbed muon site (weakly-bound case), or 
(ii)~letting them relax by $\myvec{\delta \widetilde{r}}_\mathrm{N}(\myvec{\delta x})$ to new lowest-energy positions for each $\myvec{\delta x}$ (strongly-bound case) while keeping the center of mass fixed. 
Respectively, this corresponds to: 
(i)~assuming independent muon and nuclear \zpm [\ie, a separable \muonnuclear \wavefunction, $\psi(\myvec{\delta x}, \myvec{\delta r}_\mathrm{N}) = \psi_\mu(\myvec{\delta x}) \psi_\mathrm{N}(\myvec{\delta r}_\mathrm{N})$] in the weakly-bound case, or 
(ii)~\zpm where a quantum measurement of the muon displacement $\myvec{\delta x}$ would simultaneously also determine all nuclear displacements $\myvec{\delta r}_\mathrm{N}$ [\ie, a maximally-entangled \muonnuclear \wavefunction, $\psi(\myvec{\delta x}, \myvec{\delta r}_\mathrm{N}) \propto \psi_\mu(\myvec{\delta x}) \delta(\myvec{\delta r}_\mathrm{N} - \myvec{\delta \widetilde{r}}_\mathrm{N}(\myvec{\delta x}))$, where $\delta$ is the \dirac delta function] in the strongly-bound case. 
Although, by the variational principle, the strongly-bound potential is shallower than the weakly-bound potential (which would tend to increase muon delocalization), the effective mass $m_\mathrm{eff}$ in the strongly-bound case increases above the free-muon mass $m_\mu$ due to additional movement of nuclei with the muon (which would tend to decrease muon delocalization), meaning that muon \zpm delocalization can either increase or decrease in the strongly-bound case. 
Explicitly, assuming a linear dependence of the displacement of the $i$th nucleus of mass $m_i$ on the muon displacement, \ie, $\myvec{\delta \widetilde{r}}_{\mathrm{N} i}(\myvec{\delta x}) = \mytensor{A}_i \myvec{\delta x}$ where $\mytensor{A}_i$ is a tensor, the effective mass tensor in the strongly-bound case is given by $\mytensor{m}_\mathrm{eff} = m_\mu \idtensor + \sum_i m_i \mytensor{A}_i\tran \mytensor{A}_i$ where $\idtensor$ is the identity tensor. 
In the weakly-bound case, $m_\mathrm{eff} = m_\mu$. 
Once $V_\mathrm{eff}(\myvec{\delta x})$ and $m_\mathrm{eff}$ are known, a single-particle \schrodinger equation for the muon \wavefunction $\psi_\mu(\myvec{\delta x})$, $[-\hbar^2/(2 m_\mathrm{eff})] \nabla^2 \psi_\mu + V_\mathrm{eff} \psi_\mu = (\zpeEq) \psi_\mu$, can be numerically solved to obtain the \muonnuclear \zpm, encoded in $\psi_\mu(\myvec{\delta x})$, and the corresponding \zpe. 
This inherently single-particle approximation with 3 degrees of freedom in $\myvec{\delta x}$ (due to the assumed zero or maximal \muonnuclear entanglement) cannot describe \manybody \zpm involving more than 3 degrees of freedom (\ie, cases with partial \muonnuclear entanglement). 

\begin{ruledtabular}
\begin{table*}[!htb]
\caption{%
{\bf Muon \zpmLong parameters in \aNtwo.} 
Total muon \zpeLong ($\zpeEq$) with directional contributions $\zpeEq_j$, muon \wavefunction widths ${\Delta x}_j$, and effective muon mass renormalization $m_{\mathrm{eff}, j}/m_\mu$ along the $j = \mathrm{L}$, $\mathrm{T}_1$, and $\mathrm{T}_2$ directions, in this order, for the approximations described in the text. 
\PimdLongFirst $\zpeEq_j$ was estimated as the harmonic $\zpeEq_j$ yielding the same ${\Delta x}_j$ if $m_\mathrm{eff} = m_\mu$ is assumed (marked with *).
}
\begin{center}
\renewcommand{\arraystretch}{1.2}
\begin{tabular}{@{} l r r r r @{}}
Approximation & $\zpeEq$ ($\si{\electronvolt}$) & $\zpeEq_j$ ($\si{\electronvolt}$) & ${\Delta x}_j$ ($\si{\angstrom}$) & $m_{\mathrm{eff}, j}/m_\mu$ \\
\midrule
Adiabatic (weakly-bound)   & $0.69$     & $[0.29, 0.20, 0.20]$       & $[0.16, 0.21, 0.21]$             & $1$ \\
Adiabatic (strongly-bound) & $0.008(4)$ & $[0.0021, 0.0031, 0.0028]$ & $[0.074(4), 0.093(3), 0.079(4)]$ & $[810(160), 350(40), 520(80)]$ \\
Harmonic (weakly-bound)    & $0.58$     & $[0.17, 0.20, 0.20]$       & $[0.23, 0.21, 0.21]$             & $1$ \\
\pimd                      & $0.43(11)$*    & $[0.16(11), 0.13(1), 0.14(1)]$*      & $[0.24(8), 0.27(1), 0.25(1)]$             & $1$*
\end{tabular}
\end{center}
\label{table_zpm_parameters}
\end{table*}
\end{ruledtabular}

\cref{fig_potential}a, b show the calculated weakly- and strongly-bound adiabatic effective potentials. 
Separable potentials are assumed with one eigenaxis, $\mathrm{L} \parallel [0, 1, 1]$, along the \NtwomuNtwo complex and two, $\mathrm{T}_1 \parallel [0, 1, \bar{1}]$ and $\mathrm{T}_2 \parallel [0, 0, 1]$, transverse to it, coinciding with muon $\Gamma$-point normal mode directions from \dft (\cref{fig_potential}c). 
\cref{table_zpm_parameters} lists the total muon $\zpeEq = \sum_j \zpeEq_j$ with directional contribution $\zpeEq_j$ and muon \wavefunction widths ${\Delta x}_j = \braket{{\delta x}_j^2}^{1/2}$ along the $j = \mathrm{L}$, $\mathrm{T}_1$, and $\mathrm{T}_2$ directions under both approximations. 
In the weakly-bound case, we find muon \zpm delocalization of $0.16$--$\SI{0.20}{\angstrom}$, which is significant compared to the $\SI{1.29(1)}{\angstrom}$ distance between the classical muon site and the nearest nitrogen. 
However, in the strongly-bound case we also find large directional effective muon mass renormalization 
$m_{\mathrm{eff}, j}/m_\mu$ due to strong electrostatic interactions within the \NtwomuNtwo complex and with the surrounding polaron, which leads to a large number of nitrogen nuclei following the displaced muon (up to $494 m_\mu$ in the complex and a further $494 m_\mu$ just in first shell of the polaron). 
This leads to a significantly reduced $\zpeEq$ and delocalization (\cref{table_zpm_parameters}) under the strongly-bound approximation, despite a shallower effective potential (\cref{fig_potential}a, b) compared to the weakly-bound case. 

\vspace{-1.5em} %
\subsection*{Toy model}
\vspace{-1.25em} %
Given the discrepancy between the two adiabatic approximations, the question of their applicability arises. In other words, whether \muonnuclear \zpm is minimally (weakly-bound limit), maximally (strongly-bound limit), or partially entangled. 
We address this using a toy model with the muon bound to $N = 2$ nearby effective nuclei with force constants $k_\mu$, which are further bound to a static lattice with force constants $k_\mathrm{n}$. 
If $k_\mu \ll k_\mathrm{n}$ muon displacements barely perturb the nuclei (weakly-bound limit), while if $k_\mu \gg k_\mathrm{n}$ muon displacements strongly displace nearby nuclei (strongly-bound limit). 
We estimate the ratio $k_\mu / k_\mathrm{n}$ in \aNtwo from the ratio of adiabatic potentials as $k_{\mu,j} / k_{\mathrm{n},j} = V_\mathrm{eff}^\mathrm{weak}/V_\mathrm{eff}^\mathrm{strong} - 1 \approx 7$, $12$, and $9$ along the $j = \mathrm{L}$, $\mathrm{T}_1$, and $\mathrm{T}_2$ directions, respectively (\cref{fig_potential}a, b). 
This excludes the weakly- but not the strongly-bound adiabatic approximation. 
However, the relatively large ratio $k_\mu / k_\mathrm{n}$ still competes with the tendency of light particles to partially positionally decouple from heavier particles~\cite{storchak1998quantum}, which could lead to an intermediate, partially-entangled regime where single-particle (adiabatic) approximations fail. 
This is tendency is expected to be further reinforced by the enhanced mass $m_{\mathrm{n},j}$ of effective nearest nuclei in the toy model needed to obtain the same calculated effective muon mass $m_{\mathrm{eff},j}$ under the strongly-bound adiabatic approximation both from \dft (\cref{table_zpm_parameters}) and from the toy model. 
Namely, in the toy model we need $m_{\mathrm{n},j} = (1 + k_{\mathrm{n},j} / k_{\mu,j})^2 (m_{\mathrm{eff},j} - m_\mu) / N = 4.2(9) m_\mathrm{N}$, $1.7(2) m_\mathrm{N}$, and $2.6(4) m_\mathrm{N}$ along the $j = \mathrm{L}$, $\mathrm{T}_1$, and $\mathrm{T}_2$ directions, respectively, where $m_\mathrm{N}$ is the mass of a nitrogen nucleus. 

\vspace{-1.5em} %
\subsection*{Harmonic quantum approximations}
\vspace{-1.25em} %
An alternative class of approximations are harmonic methods~\cite{boxwell1993ab,moller2013quantum,manas2021quantum}, which work in the \manybody regime of partial entanglement, but are limited to strictly harmonic \muonnuclear interactions. 
These start with a $\Gamma$-point \dft phonon calculation in the classical muon site geometry (a muon is a localized defect and thus has no $q$-space dispersion). 
The usual assumption that, since muons are lighter than nuclei, muon \zpm is fully adiabatically decoupled from the lattice (yielding 3 highest-frequency normal modes $\omega_j$ describing single-particle muon \zpm; see \cref{fig_potential}c), corresponds to the weakly-bound (zero entanglement) adiabatic limit described above, but with the additional constraint of a harmonic adiabatic potential $V_\mathrm{eff}$. 
This assumption yields directional \zpe contributions $\zpeEq_j = \hbar\omega_j/2$ and directional \zpm delocalization ${\Delta x}_j = \sqrt{\hbar/(2 m_\mu \omega_j)}$. 
The $\SI{>40}{\percent}$ discrepancy between the harmonic and anharmonic weakly-bound adiabatic values of ${\Delta x}_\mathrm{L}$ thus obtained (\cref{table_zpm_parameters}) hints at a breakdown of the harmonic approximation due to strong anharmonicity (\cref{fig_potential}a). 

Although the weakly-bound adiabatic limit is exact if $m_\mu \rightarrow 0$, it cannot reproduce \nonzero \muonnuclear entanglement. 
In fact, in \aNtwo we see strong hybridization of the $\mathrm{L}$ muon normal mode with intra-molecular vibrations of both \Ntwo molecules in the \NtwomuNtwo complex due to a finite muon mass (\cref{fig_potential}c), which implies significant \muonnuclear entanglement~\cite{srivastava1990physics,kantorovich2004quantum}. 
This is detected by projecting the top 3 (normalized) phonon normal modes onto pure muon motion, summing the squared norms of the resulting (projected) phonon eigenvectors, and subtracting the value $3$, which is expected when muon normal modes do not mix with the lattice modes (\ie, in the weakly-bound limit). 
This defines an entanglement witness~\cite{plenio2007introduction,horodecki2009quantum,horodecki2021quantum} $w_1$ as $w_1 = 0$ for zero \muonnuclear entanglement (weakly-bound adiabatic limit) and $w_1 < 0$ in the entangled, \manybody case. 
For muons in \aNtwo we obtain $w_1 = -0.14 < 0$. 
This again suggests that the weakly- (and possibly also the strongly-) bound adiabatic approximation should fail, as \muonnuclear \zpm is inherently \manybody, at least in the harmonic approximation. 
Were it not for strong anharmonicity, which invalidates the approach, such complex \zpm could still be treated by the full, \manybody harmonic method by considering the entire \supercell $\Gamma$-point phonon spectrum. 

\vspace{-1.5em} %
\subsection*{Full quantum muon description}
\vspace{-1.25em} %
Finally, we turn to numerically-exact \pimd for calculating observables from arbitrary \muonnuclear \zpm~\cite{herrero2014path,ceriotti2016nuclear,markland2018nuclear}, based on discretizing imaginary-time, $T > 0$, path integrals. 
Unlike approximate methods, \pimd works even in the anharmonic \manybody regime with partial \muonnuclear entanglement~\cite{shiga2018path}. 
Using \pimd we find that the projected muon probability density is unimodal in \aNtwo (\cref{fig_potential}a, b), confirming that the muon site is unique even for quantum muons~\cite{terrill2010ab}, with no signs of quantum tunneling. 
Thoroughly testing \pimd convergence of observables against simplified toy-model simulations, we confirm that all observables are well converged by $P = 16$--$24$ \pimd beads (imaginary-time steps), except for muon \zpm delocalization ${\Delta x}_j$, where we can correct for finite-$P$ effects \via a 
careful $P \rightarrow \infty$ extrapolation scheme (see Methods and Figs.~S3 and S4). 
We find that muon ${\Delta x}_\mathrm{L}$ is large enough (\cref{table_zpm_parameters}) that anharmonic effects become significant (\cref{fig_potential}a) and harmonic approximations fail. 
Furthermore, all ${\Delta x}_j$ are larger than in the weakly- and strongly-bound adiabatic approximations, implying partial \muonnuclear entanglement and \manybody \zpm outside of the scope of single-particle approximations, as already anticipated from $\mathrm{L}$ normal mode hybridization (\cref{fig_potential}c). 
To quantify the degree of entanglement we calculate a multivariate \pearson correlation coefficient~\cite{ichiye1991collective} 
\begin{equation}
c_i = \frac{ \braket{\myvec{\delta x}' \cdot \myvec{\delta r}_{\mathrm{N} i}'} }{ \sqrt{ \braket{| \myvec{\delta x}' |^2 } \braket{| \myvec{\delta r}_{\mathrm{N} i}' |^2 } } } \in [-1, 1] 
\label{eq_corr}
\end{equation} 
between the centered muon displacement $\myvec{\delta x}' = \myvec{\delta x} - \braket{\myvec{\delta x}}$ and the centered displacement $\myvec{\delta r}_{\mathrm{N} i}' = \myvec{\delta r}_{\mathrm{N} i} - \braket{\myvec{\delta r}_{\mathrm{N} i}}$ of a nucleus $i$, \via \pimd. 
The quantity $w_{2 i} = -|c_i| \in [-1, 0]$ is a witness of \muonnuclear entanglement~\cite{plenio2007introduction,horodecki2009quantum,horodecki2021quantum}, since $w_{2 i} = 0$ for separable (weakly-bound) states (as then $\braket{\myvec{\delta x}' \cdot \myvec{\delta r}_{\mathrm{N} i}'} = \braket{\myvec{\delta x}'} \cdot \braket{\myvec{\delta r}_{\mathrm{N} i}'} = 0$), which means that $w_{2 i} < 0$ implies entanglement. 
On the other hand, in the strongly-bound case (maximal entanglement) with $\myvec{\delta \widetilde{r}}_{\mathrm{N} i} \propto \myvec{\delta x}$ we find $|c_i| = \SI{100}{\percent}$. 
In \aNtwo the degree of \muonnuclear entanglement is $c_i \approx \SI{30(4)}{\percent}$ with the two nearest nitrogen nuclei, and decays with distance (\cref{fig_site_polaron_ent}c). 
This confirms that \muonnuclear \zpm in \aNtwo is partially entangled and thus inherently \manybody. 
We note that a \manybody harmonic approximation calculation~\cite{kantorovich2004quantum,monacelli2021stochastic} would yield a similar $c_i$ (${\approx}\SI{26}{\percent}$) between the muon and the two nearest nitrogen nuclei. 

\begin{ruledtabular}
\begin{table}[!tb]
\caption{%
{\bf Quadrupolar coupling constant of \aNtwo.} 
\NqccStaticLongFirst $\nqccStaticEq$ of \Nisotope in \aNtwo from \uncalibrated \dftLongFirst calculations, and calculations calibrated by $\nqccCaliFactEq$ obtained from fits of experimental \qlcrLongFirst spectra with predictions for a classical (\dftLong, \dft) or a quantum muon (\pimdLong, \pimd; see \cref{fig_qlcr_theory_and_fits}). 
Discrepancy from the literature value $\SI{-5.39(5)}{\mega\hertz}$~\cite{scott1976solid} is given in standard deviations $\sigma$.
}
\begin{center}
\renewcommand{\arraystretch}{1.2}
\begin{tabular}{@{} l r r r r @{}}
Calibration & $T$ ($\si{\kelvin}$) & $\nqccCaliFactEq$ & $\nqccStaticEq$ ($\si{\mega\hertz}$) & Discrepancy ($\sigma$) \\ 
\midrule
None                                  & & $1$        & $-5.76$ & $-7.4(2)$    \\
Classical \muonPlus & $1.8$  & $1.106(7)$ & $-5.21(3)$ & $3.1(6)$ \\
 & $5$    & $1.103(4)$ & $-5.22(2)$ & $3.2(7)$ \\
Quantum \muonPlus & $1.8$           & $1.076(6)$ & $-5.35(3)$ & $0.7(1)$ \\
 & $5$             & $1.073(4)$ & $-5.37(2)$ & $0.4(1)$
\end{tabular}
\end{center}
\label{table_qlcr_fits}
\end{table}
\end{ruledtabular}

\begin{figure}[!t]
\centering
\includegraphics[width=1\columnwidth]{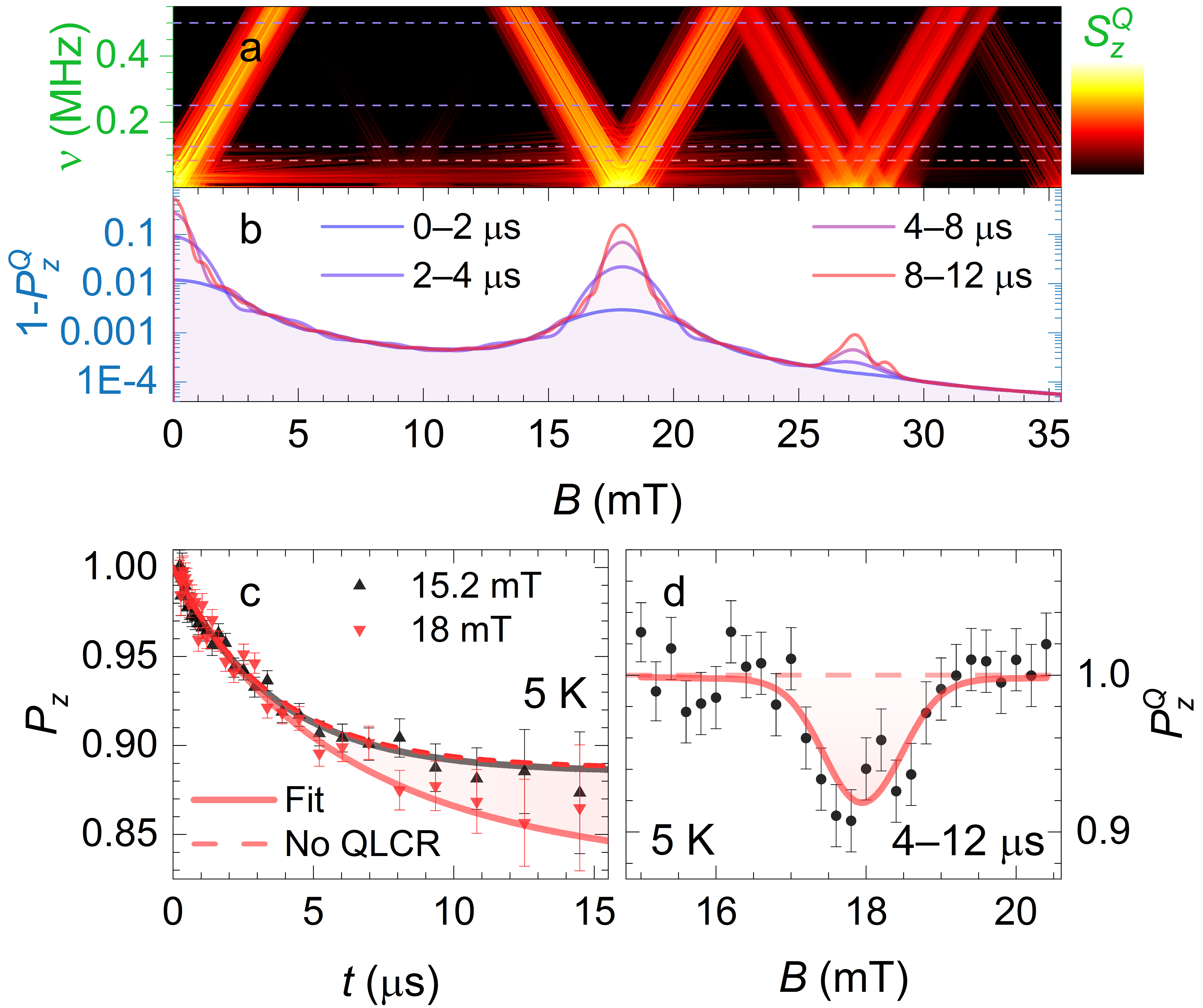}
\caption{%
{\bf Muon \qlcrLong spectra and fits.} 
{\bf a} Calibrated \fourier-transformed \qlcrLongFirst spectra $S_\mathrm{z}^\mathrm{Q}(\nu, B)$ \vs frequency $\nu$ from \pimdLongFirst calculations with $24$ beads (see Methods) for time-differential \qlcr analysis~\cite{lord2014time}. 
{\bf b} Muon-lifetime weighed integrals of $1 - P_\mathrm{z}^\mathrm{Q}(t, B)$ over different time windows for integral \qlcr analysis. 
{\bf c} Experimental \qlcr data at $\SI{5}{\kelvin}$ on ($\SI{18}{\milli\tesla}$) and off resonance ($\SI{15.2}{\milli\tesla}$) and the global fit with (solid) and without the \qlcr term (dashed). 
{\bf d} Global fit of $\SI{5}{\kelvin}$ \qlcr data with \pimd predictions integrated over a $4$--$\SI{12}{\micro\second}$ time window \vs $B$. 
Error bars on {\bf c} and {\bf d} represent a statistical uncertainty of one standard deviation. 
}
\label{fig_qlcr_theory_and_fits}
\end{figure}

\vspace{-1.5em} %
\subsection*{\qlcr measurements and \nqcc of \aNtwo}
\vspace{-1.25em} %
Detailed knowledge of the \manybody \muonnuclear \zpm afforded by \pimd allows us to derive an important constant: the \nqccStatic of \Nisotope in \aNtwo, defined as $\nqccStaticEqLongInlineFirst$, where $Q = \SI{2.044(3)}{\femto\meter\squared}$ is the quadrupolar moment of \Nisotope~\cite{tokman1997nuclear,rumble2021crc}, and $q_0$ is the largest eigenvalue of the \efgLongFirst eigenvalue at the \Nisotope position, assuming static classical nitrogen nuclei. 
The experimental \nqcc at $T = 0$, $\nqccEq = \nqccZeroTempOrderParamEq \nqccStaticEq$, is reduced from the static $\nqccStaticEq$ by the solid-nitrogen order parameter $\nqccZeroTempOrderParamEq = 0.863(8) < 1$ due to the \zpm of nitrogen nuclei~\cite{scott1976solid}. 
Using \dft we calculate $\nqccStaticEq$ by assuming classical point-particle nuclei in pristine \aNtwo, obtaining the \abinitio value $\nqccStaticEqDFT = \SI{-5.76}{\mega\hertz}$, which differs from the accepted experimental value $\nqccStaticEq = \SI{-5.39(5)}{\mega\hertz}$ by $\SI{{\approx}6.9}{\percent}$ or ${\approx}7.4$ standard deviations $\sigma$ (\cref{table_qlcr_fits}). 
The discrepancy arises from systematic errors in \dft calculations of \efgs, which are apparently overestimated by a factor $\nqccCaliFactEq = \nqccStaticEqDFT/\nqccStaticEq \approx 1.07(1)$. 
To obtain an independent estimate of 
$\nqccStaticEq = \nqccCaliFactEq^{-1} \nqccStaticEqDFT$ 
we therefore need to determine $\nqccCaliFactEq$, and calibrate our \abinitio results against experiment. 

To achieve this we performed precision \qlcr~\musr measurements on \aNtwo. 
In a \qlcr experiment muons stop near $I \geq 1$ nuclei (here \Nisotope), while a longitudinal magnetic field $B$ is swept in small steps at a given $T$ and the muon polarization $P_\mathrm{z}(t)$ is measured. 
For $B$ where the \zeeman splitting of muon spin energy levels approaches the splitting of nuclear energy levels due to \efgs at nuclear positions, resonant cross-relaxation of \muonnuclear spins occurs, resulting in a sharp dip of the measured late-time $P_\mathrm{z}(t)$~\cite{abragam1984spectrometrie,storchak1992muon,lord2014time}. 
If we also have a good description of \muonnuclear \zpm (\eg, \via \pimd), which shifts and reshapes \qlcr spectra $P_\mathrm{z}(t, B)$, we can extract muonated-sample \efgs and compare them to \abinitio \dft predictions of them to find the sought after \dft calibration factor $\nqccCaliFactEq$. 
Since our modeling showed strongly time-window-dependent widths of \qlcr spectral peaks in \aNtwo (\cref{fig_qlcr_theory_and_fits}b), we went beyond conventional integral or differential \qlcr analysis~\cite{lord2014time}, and performed a simultaneous fit to our data over all measured $B$ and $t < \SI{18}{\micro\second}$ (no binning with a $\SI{16}{\nano\second}$ time resolution), at each measured $T$, using a simple global model 
\begin{equation}
P_\mathrm{z}(t, B) = A_\mathrm{Q} P_\mathrm{z}^\mathrm{Q}(t, B) + A_\mathrm{\muoniumEq} e^{-\lambda_\muoniumEq t} + A_\mathrm{bgd}(B) 
\label{eq_qlcr_fit_model}
\end{equation} 
where $A_Q$ is the amplitude of \qlcr signal, $P_\mathrm{z}^\mathrm{Q}$ is the \abinitio simulated \qlcr signal (for either the classical or the quantum case; see Methods) at a given $\nqccCaliFactEq$, which was a fit parameter (\cref{fig_qlcr_theory_and_fits}a, b, and Supplementary Fig.~6), while $A_\mathrm{\muoniumEq}$ describes the muonium fraction, known to form in \aNtwo~\cite{storchak1999muonium} and relaxing with rate $\lambda_\muoniumEq$. 
The contribution $A_\mathrm{bgd}(B)$ due to muons stopped outside the sample was modeled as a constant plus a weak \lorentzian of $B$, with fits yielding a field variation of $\SI{<10}{\percent}$ of the \qlcr signal (\cref{fig_qlcr_theory_and_fits}c), which does not affect our conclusions. 
These fits described ${>}28,000$ experimental data points at each $T$ using only $8$ global fit parameters (\cref{fig_qlcr_theory_and_fits}c, d), with good fit quality (see Methods) and without systematic deviations at any $t$ or $B$ (Supplementary Fig.~5). 
This held true for both classical and quantum cases, as the main effect of \muonnuclear \zpm was a simple shift of the main \qlcr peak, which was absorbed into the fitted value of $\nqccCaliFactEq$. 
Extracting $\nqccCaliFactEq$ from fits of \qlcr spectra under the classical muon \ansatz we thus obtain 
$\nqccStaticEq = \nqccCaliFactEq^{-1} \nqccStaticEqDFT = \SI{-5.22(2)}{\mega\hertz}$, 
which still differs from the accepted experimental value by ${\approx}3.2 \sigma$ (\cref{table_qlcr_fits}). 
However, extracting $\nqccCaliFactEq$ from fits of \qlcr spectra using a \pimd description of \muonnuclear \zpm, which is well-converged without the need for finite-$P$ extrapolation (see Methods and Supplementary Fig.~3), we obtain excellent agreement with previous experiments, with the final value $\nqccStaticEq = \SI{-5.36(2)}{\mega\hertz}$ (\cref{table_qlcr_fits}) statistically even more accurate than the previous best estimate of $\SI{-5.39(5)}{\mega\hertz}$~\cite{scott1976solid}. 

\vspace{-1.5em} %
\section*{Discussion}
\vspace{-1.25em} %
Based on the above analysis we are able to propose several rules of thumb for when to expect quantum muon effects, which have the potential to significantly impact the interpretation of \musr experiments. 
\Manybody quantum muon effects are expected to be large whenever: 
(i)~the muon is strongly (chemically) bound to the crystal lattice (see Toy model), \eg, when the material is ionic or contains very electronegative atoms, like \ch{F-}, \ch{Cl-}, \ch{Br-}, \ch{O^{2-}}, or \ch{N^{3-}}, or functional groups, like \ch{OH-}, that strongly attract the muon, or 
(ii)~when the phonon normal modes of the pristine material are high-frequency, allowing them to hybridize with muon normal modes (\cref{fig_potential}c). 
For example, the latter scenario is expected when the local chemical bonds in the material around the muon site are strong, \eg, ionic or strong covalent (\eg, double or triple) bonds, or when the material contains light atoms; especially if the bonds are weaker further away from the muon site, like in molecular crystals, as discussed in the Toy model section. 
Under the same conditions, muon \zpe is expected to be high, which could affect the energy ordering, and thus the expected occupancy, of candidate muon sites. 
These same conditions are also expected to lead to strong deformations of the crystal structure around the muon sites, which suggests a simple rule: if the muon strongly perturbs the crystal structure classically, it likely also becomes quantum entangled with it with a high \zpe. 
On the other hand, anharmonic quantum muon effects are expected to be important when: 
(i)~the muon is highly delocalized, which occurs when it is weakly bound to neighboring atoms, \eg, at interstitial sites, or, alternatively, 
(ii)~if the effective muon potential is inherently highly anharmonic (\cref{fig_potential}a). 
Muon delocalization and anharmonicity can strongly affect the interpretation of \musr results, since they directly affect the nature and strength of the coupling of the muon to the local magnetic fields. 
When either just many-body or anharmonic effects are present, they can be treated using appropriate harmonic or adiabatic approximations, respectively. 
However, as seen in the case of muons in \aNtwo, \manybody and anharmonic effects can also be present at the same time requiring their careful examination using more general quantum methods. 
Examples of these include numerically-exact methods like \pimd or the recently-proposed \miwLongAdj approach~\cite{hall2014quantum,simone2018computational}. 
We note that the discussion above is independent of the local symmetry, or any lack thereof, at the muon site. 

Beyond muons, our unified description of light-particle \zpm, where quantum regimes with well-defined approximations arise in certain limits of particle--lattice entanglement and anharmonicity, directly applies also to other light particles and light-atom nuclei, \eg, hydrogen and lithium, in solids. 
Such an extension is of immense interest, as nuclear quantum effects were shown to play a crucial role, for example, in stabilizing recent record high-$T_\mathrm{c}$ hydride superconductors \HThreeS~\cite{drozdov2015conventional,errea2015hydrogen,errea2016quantum} and \LaHTen~\cite{drozdov2019superconductivity,somayazulu2019evidence,errea2020quantum}, as well as the contentious supposed room-temperature superconductors \LuHThreeN~\cite{dasenbrock2023evidence,ming2023absence} and \CSHEight~\cite{snider2020room,snider2022retraction,hirsch2021unusual,eremets2022high}, and in explaining their huge isotope effects, as calculated \via the self-consistent harmonic approximation~\cite{errea2015hydrogen,errea2016quantum,errea2020quantum}. 
A similar situation arises in the recent record-density hydrogen storage material \hydrogenStorageCHCompound~\cite{ranieri2022formation}, which was found to be stabilized by nuclear quantum effects \via the harmonic approximation, and in calculating solvation free energies of Li and F~\cite{duignan2017real} \via the quasi-harmonic approximation. 
In all of these materials, a careful examination of the underlying entanglement and anharmonicity regimes of nuclear \zpm using the approaches described in this paper could reveal any possible limitations on the validity of the approximations used and provide a systematic way of finding even more accurate nuclear \zpm descriptions. 
In this way, our understanding of these highly intriguing quantum materials could be substantially improved. 

In conclusion, we have performed an analysis of \muonnuclear \zpm in \aNtwo, culminating in a precision determination of the static \nqcc of \Nisotope in pristine \aNtwo using complementary \qlcr experiments and state-of-the-art \abinitio~\dftpimd calculations, significantly improving the accuracy of this constant over the previously known value. 
We also discovered an electric-dipole polaron with strong effective mass renormalization around muons in \aNtwo, which might affect the interpretation of experiments purportedly showing quantum tunneling of muonium in this material~\cite{storchak1998quantum}. 
In fact, a similar polaron might generically be expected in molecular crystals with charged impurities. 
More broadly, our work demonstrates the need to consider quantum effects when interpreting \musr data, paying attention to the presence of \muonnuclear entanglement and anharmonicity, to guide the choice of the applicable \zpm approximation. 
This unified perspective on light-particle \zpm is generally applicable and should be explored further also in the context of material science and quantum chemistry, where it promises to offer a way of finding accurate and computationally-efficient descriptions of nuclear quantum effects of light atoms across a wide range of quantum materials. 

{\begingroup 
\fontsize{8.5pt}{10pt}\selectfont 
%
\vspace{-1.5em} %
\section*{Methods}
\vspace{-1.25em} %
%

\subsection*{\dft calculations}
\vspace{-1.25em} %
Both the pristine and muonated low-$T$ low-pressure $\alpha$ phase of solid nitrogen, \aNtwo, was studied using the \castep plane-wave \abinitio \dftLongFirst code~\cite{clark2005first} using the \pbe \xcFuncLong~\cite{perdew1996generalized} and ultrasoft pseudopotentials. 
Calculations were carried out on a \twoCubed \supercell to avoid finite-size effects around the implanted muon, while a $\SI{1200}{\electronvolt}$ \planewaveAdj energy cutoff and \twoCubed \monkpack grid~\cite{monkhorst1976special} reciprocal-space sampling was chosen to achieve numerical convergence. 
A neutral cell was used for pristine structure calculations, while a positive elementary charge per \supercell was applied in calculations of muonated structures, to account for the positive charge of the muon. 
All calculations were converged to within a total energy tolerance of \SI{0.1}{\nano\electronvolt\per\mathrm{atom}} in the \scfLongFirst \dft loop, while geometry relaxation tasks were converged to within a tight \SI{5}{\milli\electronvolt\per\angstrom} force tolerance on the muon and nuclei. 
Furthermore, to properly account for weak cohesive \vdwLongFirst dispersion forces between \Ntwo molecules, which are usually underestimated in pure \dft, a \tsLongFirst semi-empirical dispersion correction scheme was applied in a \dftd approach~\cite{tkatchenko2009accurate}. 
We note, though, that the results of ordinary \dft with an \adhoc isotropic external hydrostatic pressure of $\SI{0.5}{\giga\pascal}$ (chosen to reproduce the experimental zero-pressure unit cell volume of pristine \aNtwo) were practically indistinguishable from the results of full \dftd calculations, even in the presence of an implanted muon. 
This indicates that in \aNtwo the dominant \vdw dispersion force contribution is a simple isotropic attraction among all nuclei and the muon. 

Despite competing suggestions of a \PaThreeFirst or a \PTwoOneThreeFirst cubic crystallographic structure of pristine \aNtwo in the literature~\cite{scott1976solid}, we find that the higher-symmetry \PaThree structure is moderately preferred in our \dft calculations (by $\SI{\sim 0.05}{\electronvolt}$ per unit cell), in line with recent consensus~\cite{erba2011post,maynard2020re,rumble2021crc}. 
In this structure the centers of \Ntwo molecules form a \fccLong lattice (Supplementary Fig.~1a). 

\vspace{-1.5em} 
\subsubsection*{Classical muon stopping site}
\vspace{-1.25em} %
To find candidate muon stopping sites in \aNtwo a muon's initial position was randomly seeded in the unit cell and the full crystal geometry (including the muon position) relaxed at a fixed, experimental cell volume until convergence. 
This process was repeated ${>}20$ times to generate a list of candidate muon stopping sites, which were then grouped into clusters by identifying those candidate muon sites within $\SI{{<}0.7}{\angstrom}$ of each other (or of any in-between sites) as belonging to the same cluster. 
Here the distance between two arbitrary candidate muon sites $\myvec{r}_{\mu 1}$ and $\myvec{r}_{\mu 2}$ was taken as the minimal possible real-space distance $\abs{\myvec{r}_{\mu 1} - T \myvec{r}_{\mu 2}}$ under any symmetry operation $T$ in the pristine \aNtwo crystallographic space group \PaThree (including discrete translational, rotational and reflection symmetries). 
This is because the implanted muon is the sole source of local symmetry breaking in the crystal and thus muon sites related by pristine space-group symmetry operations should be regarded as identical. 
A muon site cluster can thus be thought of as a connected component in a graph whose vertices are the calculated muon sites that are considered adjacent if their minimal, symmetry-reduced distance is below a chosen threshold ($\SI{0.7}{\angstrom}$ in our case). 
We note that this symmetry-aware clustering algorithm based on graph theory is also implemented in the user-friendly \mufinder program for determining and analysing muon stopping sites~\cite{huddart2022mufinder,huddart2020muon}. 
In the end, we find a single muon site cluster in \aNtwo, which lies at the point midway between the centers of the two neighboring \Ntwo molecules of the pristine \aNtwo structure (Supplementary Fig.~1b). 

$\Gamma$-point phonon spectra were calculated on a \twoCubed \supercell of \aNtwo, with the whole isotope effect plot (\cref{fig_potential}c) generated from a single \dft calculation of phonon normal modes and frequencies. 
This was achieved by first reconstructing the \dmatLongFirst~\cite{srivastava1990physics,kantorovich2004quantum} of the muonated crystal, reweighting it by the desired muon isotope mass $m_\mu$, and then rediagonalizing it to obtain the new phonon normal modes and frequencies. 

\vspace{-1.5em} 
\subsubsection*{Choice of the \xcFuncLong}
\vspace{-1.25em} %
We note that switching to an \apriori less accurate \ldaLongFirst \dft \xcFuncLong~\cite{clark2005first} does not significantly alter our results. 
Namely, the classical \muonnuclear distance in the \NtwomuNtwo complex changes by just $-\SI{0.01}{\angstrom}$, the total harmonic \zpe of the 3 highest-frequency (muon) normal modes changes by $+\SI{4.1}{\percent}$ compared to the \pbe functional, and the harmonic (under)estimate of the entanglement coefficient $c_i$ of the muon with the two nearest nitrogen nuclei changes from ${\approx}\SI{26}{\percent}$ to ${\approx}\SI{24}{\percent}$. 
The anharmonicity measure $V_\mathrm{eff}^\mathrm{weak}(2 \myvec{\delta x})/(4 V_\mathrm{eff}^\mathrm{weak}(\myvec{\delta x}))$, which would equal $1$ for a purely harmonic potential, at a typical displacement $2 \delta x = +\SI{0.24}{\angstrom}$ along the most anharmonic direction $\mathrm{L}$ (\cref{fig_potential}a) changes slightly, from $1.8$ to $1.6$, when switching from \pbe to \lda. 
\Manybody and anharmonic \zpm effects of muons in \aNtwo are thus rather robust against the choice of the \xcFuncLong. 
Nevertheless, we used \pbe for all the results reported in the manuscript, as described above, since it is in general expected to be much more accurate than \lda for describing electronic systems, such as \aNtwo, where electrons are not highly correlated. 

\vspace{-1.5em} 
\subsection*{Classical \md and quantum \pimd calculations}
\vspace{-1.25em} %
For \pimd calculations an NVT statistical ensemble of up to $P = 32$ beads was simulated for up to $S = 8,600$ steps of $\SI{0.5}{\femto\second}$ with a $T = \SI{20}{\kelvin}$ stochastic \langevin thermostat. 
Satisfactory convergence in bead number was achieved already for $P = 16$--$24$ beads for all \qlcr resonance parameters (Supplementary Fig.~3) and positional observables (Supplementary Fig.~4) except for the muon \wavefunction widths (Supplementary Fig.~4a), which had to be extrapolated to the $P \rightarrow \infty$ limit (see heading Extrapolation of \pimd muon widths). 
Classical \md simulations (which can be interpreted as $P = 1$ \pimd simulations) used the same statistical ensemble, time step, temperature, and thermostat and were run for up to $S = 12,000$ steps. 
Care was taken to ensure proper thermalization of the ensembles before observables were extracted from them. 

Both \pimd and \md simulations produce a list of $S P$ \muonnuclear configurations $\{ \myvec{r}_i^{(s,p)} \}_{i=1}^K$, where $s = 1, \dots, S$ is the time step, $p = 1, \dots, P$ is the bead index, and $\myvec{r}_i$ is the position of one muon or nucleus $i$ out of $K$ muons and nuclei present in the system. %
These \muonnuclear configurations follow the corresponding \boltzmann statistical probability distribution for finding muons and nuclei at these positions. 
In the case of \md this is the $T > 0$ classical thermal probability distribution $\rho_\mathrm{c}(\myvec{r}_1, \dots, \myvec{r}_K )$, while in \pimd this is the quantum thermal probability distribution $\rho_\mathrm{q}(\myvec{r}_1, \dots, \myvec{r}_K)$, which at $T = 0$ would coincide with the ground-state \wavefunction's $\ket{\psi}$ probability distribution $\rho_\mathrm{q}(\myvec{r}_1, \dots, \myvec{r}_K) = |\psi(\myvec{r}_1, \dots, \myvec{r}_K)|^2$. 
The thermal expectation value $\braket{O}$ of any observable $O$ that depends only on \muonnuclear positions can thus be approximated from numerical \pimd or \md samples by calculating the average 
\begin{equation}
\braket{O} \approx \frac{1}{S P} \sum_{s=1}^S \sum_{p=1}^P O(\myvec{r}_1^{(s,p)}, \dots, \myvec{r}_K^{(s,p)}) 
\label{eq_pimd_md_expect}
\end{equation}
For observables that were expensive to calculate [\eg, the \efgsLongFirst, which require a full \dft calculation for each \muonnuclear configuration in the above average] a further \montecarloAdj approximation to the average was employed by randomly sampling only a small number of $s$ and $p$ (${\approx}120$ in the case of \efgs) in \cref{eq_pimd_md_expect}. 
For certain numerically-noisy observables, like the directional \wavefunction widths in {\cref{table_zpm_parameters}}, multiple runs at a given $P$ were merged together to improve statistics and provide a more reliable numerical estimate. 

\vspace{-1.5em} 
\subsubsection*{\zpm and quantum entanglement from \pimd}
\vspace{-1.25em} %
A numerical estimate of the average position $\braket{\myvec{r}_i}$ of a given muon or nucleus $i$ can be calculated in this way by choosing $O(\myvec{r}_1, \dots, \myvec{r}_K) = \myvec{r}_i$ in \cref{eq_pimd_md_expect}. 
From this we can estimate the squared gyration radius ${\Delta r}_i^2 = \braket{|\myvec{\delta r}_i|^2} = \braket{|\myvec{r}_i - \braket{\myvec{r}_i}|^2}$, which corresponds to choosing $O = |\myvec{r}_i - \braket{\myvec{r}_i}|^2$, and the covariance $\braket{\myvec{\delta r}_i \cdot \myvec{\delta r}_{i'}}$ by choosing $O = (\myvec{r}_i - \braket{\myvec{r}_i}) \cdot (\myvec{r}_{i'} - \braket{\myvec{r}_{i'}})$. 
Using these we can then extract a numerical estimate of the multivariate \pearson correlation coefficient $c_{i i'} = \braket{\myvec{\delta r}_i \cdot \myvec{\delta r}_{i'}}/({\Delta r}_i {\Delta r}_{i'}) \in [-1, 1]$~\cite{ichiye1991collective}. If this is \nonzero as $T \rightarrow 0$ it implies ground-state positional entanglement of muons or nuclei $i$ and $i'$ (see Results). 
We can also calculate the squared \zpm delocalization of a muon ${\Delta x}_j^2$ along a given direction $j = \mathrm{L}, \mathrm{T}_1, \mathrm{T}_2$, by choosing $O = |(\myvec{\Delta x} - \braket{\myvec{\Delta x}}) \cdot \myvec{\hat{v}}_j|^2$ for the corresponding unit vector $\myvec{\hat{v}}_j$ along this direction. 

\vspace{-1.5em} 
\subsubsection*{Extrapolation of \pimd muon widths}
\vspace{-1.25em} %
While the covariance of \muonnuclear positions (Supplementary Fig.~4b), the \wavefunction widths of the nitrogen nuclei (Supplementary Fig.~4c), and \qlcr resonance parameters (Supplementary Fig.~3) all fully converge for $P = 16$--$24$ beads, the muon \wavefunction widths $\Delta x_j$ do not (Supplementary Fig.~4a). 
However, numerical estimates of expected \pimd convergence under the harmonic toy model described in the main text indicate that the muon \wavefunction width is expected to be underestimated by the same factor of $g \approx 0.55(2)$ at $P = 24$ compared to the $P \rightarrow \infty$ limit in all three directions $j$ (Supplementary Fig.~4a). 
This allows us to estimate the true values of ${\Delta x}_j$ in the $P \rightarrow \infty$ limit (\cref{table_zpm_parameters}) by dividing the raw \pimd values ${\Delta x}_j = 0.13$, $0.15$, and $\SI{0.14}{\angstrom}$ along the $\mathrm{L}$, $\mathrm{T}_1$, and $\mathrm{T}_2$ directions, respectively, by this factor $g$. 
Moreover, as these widths feature in the multivariate \pearson correlation coefficient $c_i$ in \cref{eq_corr}, we estimate the true values of this entanglement proxy $c_i$ in the $P \rightarrow \infty$ limit (\cref{fig_site_polaron_ent}c) by multiplying raw \pimd values by the same factor $g$. 

As this procedure mostly neglects anharmonic effects, which are relevant along the $j = \mathrm{L}$ direction, we estimate the additional relative systematic uncertainty due to this omission along the direction $j = \mathrm{L}$ from the relative difference of ${\Delta x}_j$ values in the anharmonic weakly-bound adiabatic approximation and the harmonic approximation (\cref{table_zpm_parameters}), as a proxy for the influence of anharmonic effects. 
This yields an additional relative systematic uncertainty of at most ${\approx}\SI{30}{\percent}$ in the muon \wavefunction width along the $j = \mathrm{L}$ direction, and ${\approx}\SI{10}{\percent}$ in the entanglement witness $c_i$. 

Finally, we note the robustness of this extrapolation scheme, as even the harmonic weakly-bound adiabatic approximation (corresponding to $N = 0$ dynamical nuclei around the muon in the toy model), which is expected to be much less accurate than the $N = 2$ toy model for estimating \pimd convergence, yields a similar estimate of $g \approx 0.47(2)$ at $P = 24$. 

\vspace{-1.5em} 
\subsubsection*{Thermal effects and classical \md}
\vspace{-1.25em} %
Since \pimd calculations require a finite temperature for numerical convergence ($T = \SI{20}{\kelvin}$ in our calculations) we also checked that the observed quantum effects were not masked by thermal excitations by comparing quantum \pimd calculations against complementary classical \md simulations at the same $T$. 
We find that the classical thermal spread of \muonnuclear positions is indeed predicted to be much smaller than the quantum \wavefunction widths (\cref{table_zpm_parameters}), giving just ${\Delta x}_j = 0.08$ and $\SI{0.07}{\angstrom}$ along $j = \mathrm{L}$ and $\mathrm{T}_{1,2}$, respectively, as can be expected from the steepness of the calculated adiabatic and harmonic effective potentials (\cref{fig_potential}a, b) in relation to the thermal equipartition energy $\kB T/2 = \SI{1.7}{\milli\electronvolt}$ at this $T$, where $\kB$ is the \kBLong. 

Furthermore, quantum-mechanically a transition from $T$-independent ground-state \zpm to thermally-excited, $T$-dependent quantum behavior along a given direction $j$ upon increasing $T$ is only expected when $\kB T$ becomes comparable to the splitting $\Delta E_j$ between the ground-state and the first excited \zpm state. 
Namely, when $\kB T \ll \Delta E_j$ the system is effectively frozen in its quantum ground state,
in contrast to the classical point-particle prediction of a $T$-dependent spread of \muonnuclear positions down to the lowest $T$. 
Under the harmonic approximation this splitting can be obtained as $\Delta E_j = 2\zpeEq_j$, while $\Delta E_j > 2\zpeEq_j$ when anharmonicity that increases the effective potential at larger displacements is present, as, \eg, in the case of muons in \aNtwo along the $j = \mathrm{L}$ direction (\cref{fig_potential}a). 
From \cref{table_zpm_parameters} we can see that all considered approximations point to muons in \aNtwo being in the low-$T$, $\kB T \ll \Delta E_j$ regime along all directions $j$ at the considered temperatures. 
We thus conclude that thermal effects are irrelevant for muon \zpm in \aNtwo at these low $T$, and that the \muonnuclear system is, in fact, in its $T$-independent quantum ground state, at least locally around the muon. 

\vspace{-1.5em} 
\subsection*{\qlcr spectra calculations}
\vspace{-1.25em} %
If we were to assume fixed classical \muonnuclear positions $\{ \myvec{r}_i \}_{i=1}^K$ and the corresponding \efg tensors $\{ \mytensor{V}_i \}_{i=1}^K$ at those positions (these can be calculated from the \muonnuclear positions using \dft), the muon and nuclear spins would interact \via a \hamiltonian $\hamiltonianEq$ composed of a \zeeman contribution $\hamiltonianEq_\mathrm{Z}$, a local quadrupolar contribution $\hamiltonianEq_\mathrm{Q}$, and a dipole coupling contribution $\hamiltonianEq_\mathrm{D}$~\cite{abragam1984spectrometrie,storchak1992muon,ashbrook2006structural,slichter1990principles} 
\begin{equation}
\begin{split}
\hamiltonianEq &= \hamiltonianEq_\mathrm{Z} + \hamiltonianEq_\mathrm{Q} + \hamiltonianEq_\mathrm{D} \\
\hamiltonianEq_\mathrm{Z} &= -\sum_i \gamma_i \hbar \myvec{S}_i \cdot \myvec{B} \\
\hamiltonianEq_\mathrm{Q} &= \sum_i \frac{e_0 Q_i}{2 S_i (2 S_i - 1)} \myvec{S}_i \cdot \mytensor{V}_i \myvec{S}_i \\
\hamiltonianEq_\mathrm{D} &= \sum_{\braket{i i'}} \frac{\mu_0 \gamma_i \gamma_{i'} \hbar^2}{4 \mathrm{\pi} | \myvec{r}_{i i'} |^3} \left[ \myvec{S}_i \cdot \myvec{S}_{i'} - 3 (\myvec{S}_i \cdot \normvec{r}_{i i'}) (\myvec{\hat{r}}_{i i'} \cdot \myvec{S}_{i'}) \right] 
\end{split}
\label{eq_qlcr_ham}
\end{equation}
where $\gamma_i$ is the gyromagnetic ratio of muon or nucleus $i$ with spin vector $\myvec{S}_i$, spin size $S_i$, and nuclear quadrupole moment $Q_i$, $\myvec{B}$ is the applied magnetic field, $e_0$ is the elementary charge, $\mu_0$ is vacuum permeability, the sum in $\hamiltonianEq_\mathrm{D}$ is over unique pairs of muon or nuclei $\braket{i i'}$, $\myvec{r}_{i i'} = \myvec{r}_i - \myvec{r}_{i'}$ is a vector from a muon or nucleus $i'$ to $i$, and $\normvec{r}_{i i'} = \myvec{r}_{i i'}/|\myvec{r}_{i i'}|$ is a unit vector in the same direction. 
However, due to both quantum \zpm and thermal movement the muons and nuclei described by this \hamiltonian do not have fixed classical (point-particle) positions, nor are the \efgs independent of those positions. 

To reduce the complexity of solving the full, coupled spin--positional \hamiltonian of \cref{eq_qlcr_ham}, we assume that there is no significant entanglement between the spin and spatial degrees of freedom in the $T \approx 0$ ground-state \wavefunction of the system, \ie, that $\ket{\psi} = \ket{\psi}_\mathrm{spin} \otimes \ket{\psi}_\mathrm{position}$, due to a separation of timescales for positional and spin dynamics. 
In this way we can construct an effective spin-only \hamiltonian operator by averaging out the positional degrees of freedom with a partial trace 
\begin{equation}
\hamiltonianEq_\mathrm{eff,spin} = \tr_\mathrm{position}\left\{ \rho \hamiltonianEq \right\} 
\label{eq_qlcr_eff_ham}
\end{equation} 
where $\rho$ is the total density matrix, which at $T = 0$ equals simply $\rho = \ket{\psi} \bra{\psi}$; \ie, at $T = 0$ we have $\hamiltonianEq_\mathrm{eff,spin} = {}_\mathrm{position}\!\braket{\psi | \hamiltonianEq | \psi}\!{}_\mathrm{position}$. 
In \pimd and \md simulations this can be done numerically by calculating an average over positional degrees of freedom \via \cref{eq_pimd_md_expect} where the observable $O$ is taken to be the full \hamiltonian $\hamiltonianEq$ from \cref{eq_qlcr_ham}, and at each \muonnuclear configuration sample $\{ \myvec{r}_i^{(s,p)} \}_{i=1}^K$ we recalculate the corresponding \efg tensors $V_i$ \via \dft. 
As mentioned previously, due to the time complexity of calculating the \efg tensors, the full average in \cref{eq_pimd_md_expect} is approximated \via \montecarlo sampling of ${\approx}120$ random $s$ and $p$ indices from the \pimd or \md \muonnuclear configurations $\{ \myvec{r}_i^{(s,p)} \}_{i=1}^K$. 

The numerical calculation of a single \hamiltonian sample [\cref{eq_qlcr_ham}] from the \muonnuclear configuration and the corresponding \efg tensors was performed using the \calcalc program~\cite{calcalc,berlie2022muon} by considering only the four \Nisotope nuclei in the \NtwomuNtwo complex, \ie, the four nitrogen nuclei closest to the muon, since further-away nitrogen nuclei are only very weakly dipolarly coupled to the muon and thus do not affect its relaxation much. 
A \Nisotope nuclear quadrupole moment of $Q = \SI{2.044(3)}{\femto\meter\squared}$ was used~\cite{tokman1997nuclear,rumble2021crc}. 
Once the effective spin-only \hamiltonian [\cref{eq_qlcr_eff_ham}] was thus calculated via \cref{eq_pimd_md_expect}, a \calcalc-inspired Python program was used to calculate the time-dependent muon relaxation signal $P_\mathrm{z}^\mathrm{Q}(t)$ due to \qlcr \via exact diagonalization as in \mreference\cite{lord2000muon} for a range of applied magnetic fields $B$. 
At this stage, the quadrupolar contribution $\hamiltonianEq_\mathrm{Q}$ was multiplied by an \efg calibration factor $\nqccCaliFactEq^{-1} \approx 1$, removing most of the otherwise unavoidable systematic errors of \dft (see the Results section). 
In global fits of experimental \qlcr spectra $\nqccCaliFactEq$ was adjusted in a loop until convergence to a minimal \chiSq was achieved. 
We note that in these fits no systematic deviations at any $t$ or $B$ were observed (Supplementary Fig.~5). 
The total global fit quality was $\reducedChiSqEq = 1.06$ and $1.08$ at $T = 1.8$ and $\SI{5}{\kelvin}$, respectively. 

The final, calibrated \qlcr spectra $1 - P_\mathrm{z}^\mathrm{Q}(t, B)$ exhibit a \nontrivial dependence on time $t$ and applied field $B$ and are shown in Supplementary Fig.~6. 
\cref{fig_qlcr_theory_and_fits}a shows their \fourier transforms 
\begin{equation}
S_\mathrm{z}^\mathrm{Q}(\nu, B) = \frac{1}{2 \mathrm{\pi}} \int_{-\infty}^\infty \left[ 1 - P_\mathrm{z}^\mathrm{Q}(t, B) \right] e^{-2\mathrm{\pi} \mathrm{i} \nu t} \dd t 
\label{eq_qlcr_fourier}
\end{equation}
under the convention $P_\mathrm{z}^\mathrm{Q}(-t, B) = P_\mathrm{z}^\mathrm{Q}(t, B)$ and where $\nu$ is the frequency, while \cref{fig_qlcr_theory_and_fits}b shows their weighted time integrals 
\begin{equation}
\overline{P}_\mathrm{z}^\mathrm{Q}(B) = \frac{\int_{t_1}^{t_2} P_\mathrm{z}^\mathrm{Q}(t, B) e^{-t / \tau_\mu} \dd t}{\int_{t_1}^{t_2} e^{-t / \tau_\mu} \dd t} 
\label{eq_qlcr_weighed_int}
\end{equation}
over select time windows $t \in \left[ t_1, t_2 \right]$, weighed by the mean muon lifetime $\tau_\mu = \SI{2.197}{\micro\second}$~\cite{zyla2020review} to account for the exponential decay of muons and their subsequent \poissonian counting statistics in \musr measurements~\cite{blundell2021muon}. 

\vspace{-1.5em} 
\subsection*{\qlcr measurements}
\vspace{-1.25em} %
\qlcr~\musr measurements on \aNtwo were performed on the EMU beamline~\cite{giblin2014optimising} (ISIS pulsed muon source), using a custom-built \ch{TiZr} gas condensation cell. 
A $\SI{100}{\micro\meter}$ thick \ch{Ti} window 
allowed surface muons to enter the sample volume, while a separate experiment confirmed negligible muon depolarisation in \ch{TiZr} over the studied $T$. 
A \ch{^4He} exchange gas cryostat around the cell provided control over sample $T$ down to $\SI{\sim 1.5}{\kelvin}$, while a $\sfrac{1}{8}\si{\inchQ}$ capillary connecting the cell to an external gas supply was heated along its length to avoid blockages. 
The sample was condensed from high purity (Grade 6.0) \Ntwo gas, with $\SI{\approx 1.8}{\bar\per\litre}$ of gas being condensed at $\SI{\approx 65}{\kelvin}$ to ensure the $\SI{\approx 1.6}{\centi\meter\cubed}$ sample volume was full. 
Gas pressure was maintained well above the \Ntwo triple point ($\SI{\approx 0.125}{\bar}$) throughout sample condensation to avoid deposition of the gas directly into the solid phase. 
Once condensation was complete, the sample was cooled through the freezing point at $\SI{\approx 0.6}{\kelvin/\minute}$, with the form of \musr spectra recorded at \SI{58}{\kelvin} (\bNtwo phase) compared to earlier data on this system~\cite{storchak1999muonium}, as well as our higher-$T$ data, to confirm the sample was frozen. 
%
\vspace{-1.5em} %
\section*{Data Availability}
\vspace{-1.25em} %
The data presented in this paper~\cite{datacode} are available at \url{https://dx.doi.org/10.6084/m9.figshare.23203037}.
All other data are available from the corresponding author on reasonable request.

Supplementary Movies 1 and 2 present animated versions of \cref{fig_site_polaron_ent}a and \cref{fig_site_polaron_ent}b, respectively.
%
\vspace{-1.5em} %
\section*{Code Availability}
\vspace{-1.25em} %
The computer code used to generate and/or analyse the data in this paper~\cite{datacode} is available at \url{https://dx.doi.org/10.6084/m9.figshare.23203037}. 
\endgroup} 

%

\vspace{-1.5em} 
\section*{Acknowledgments}
\vspace{-1.25em} %
The authors acknowledge helpful discussions with A. Zorko. 
Part of this work was performed at the STFC-ISIS facility. 
Computing resources were provided by the STFC Scientific Computing Department’s SCARF cluster and the Durham HPC Hamilton cluster. 
M.G., T.L., S.J.C., and F.L.P. are grateful to Engineering and Physical Sciences Research Council (EPSRC, UK) for financial support through Grants No. EP/N024028/1 and EP/N024486/1. 
M.G. is grateful to the Slovenian Research Agency (ARRS) for financial support through Projects No. Z1-1852 and J1-2461 and Programme No. P1-0125. 

\vspace{-1.5em} 
\section*{Author contributions}
\vspace{-1.25em} %
The project was originally conceived by T.L. and F.L.P. 
M.G. carried out \dft, \md, and \pimd calculations under the supervision of T.L. and S.J.C., and developed the unified description of \muonnuclear entanglement and anharmonicity. 
F.L.P. and S.P.C. conceived of and set up the \qlcr experiments and performed them with assistance from M.G. 
\qlcr data was analyzed by F.L.P., S.P.C., and M.G. with \qlcr calculations performed by M.G. and F.L.P. 
All authors discussed the results. 
M.G. wrote the paper with feedback from all the authors. 

\vspace{-1.5em} 
\section*{Competing interests}
\vspace{-1.25em} %
The authors declare no competing interests.

\vspace{-1.5em} 
\section*{Additional Information}
\vspace{-1.25em} %
%
Correspondence and requests for materials should be addressed to M.G.


\end{document}


\selectlanguage{english}

\preprint{APS/123-QED}

\title{Supplementary Information:\\\ManyBody Quantum Muon Effects and Quadrupolar Coupling in Solids}

\author{Matja\v{z} Gomil\v{s}ek\,\orcidlink{0000-0002-9152-8905}}
\email[Corresponding author. Email: ]{matjaz.gomilsek@ijs.si} 
\affiliation{Jo\v{z}ef Stefan Institute, Jamova c.~39, SI-1000 Ljubljana, Slovenia}
\affiliation{Faculty of Mathematics and Physics, University of Ljubljana, Jadranska u. 19, SI-1000 Ljubljana, Slovenia}
\affiliation{Department of Physics, Durham University, South Road, Durham DH1 3LE, United Kingdom}
\author{Francis L. Pratt\,\orcidlink{0000-0002-5919-3885}}
\affiliation{ISIS Muon Group, Science and Technology Facilities Council (STFC), Didcot OX11 0QX, United Kingdom}
\author{Stephen P. Cottrell\,\orcidlink{0000-0002-8021-6607}}
\affiliation{ISIS Muon Group, Science and Technology Facilities Council (STFC), Didcot OX11 0QX, United Kingdom}
\author{Stewart J. Clark\,\orcidlink{0000-0003-4792-7738}}
\affiliation{Department of Physics, Durham University, South Road, Durham DH1 3LE, United Kingdom}
\author{Tom Lancaster\,\orcidlink{0000-0002-6714-4215}}
\affiliation{Department of Physics, Durham University, South Road, Durham DH1 3LE, United Kingdom}

\date{\today}

\maketitle


\renewcommand{\figurename}{\bf Supplementary Fig.}

\crefformat{figure}{#2Supplementary Fig.~#1#3}


~
%
\begin{figure}[H]
\centering
\includegraphics[width=1\columnwidth]{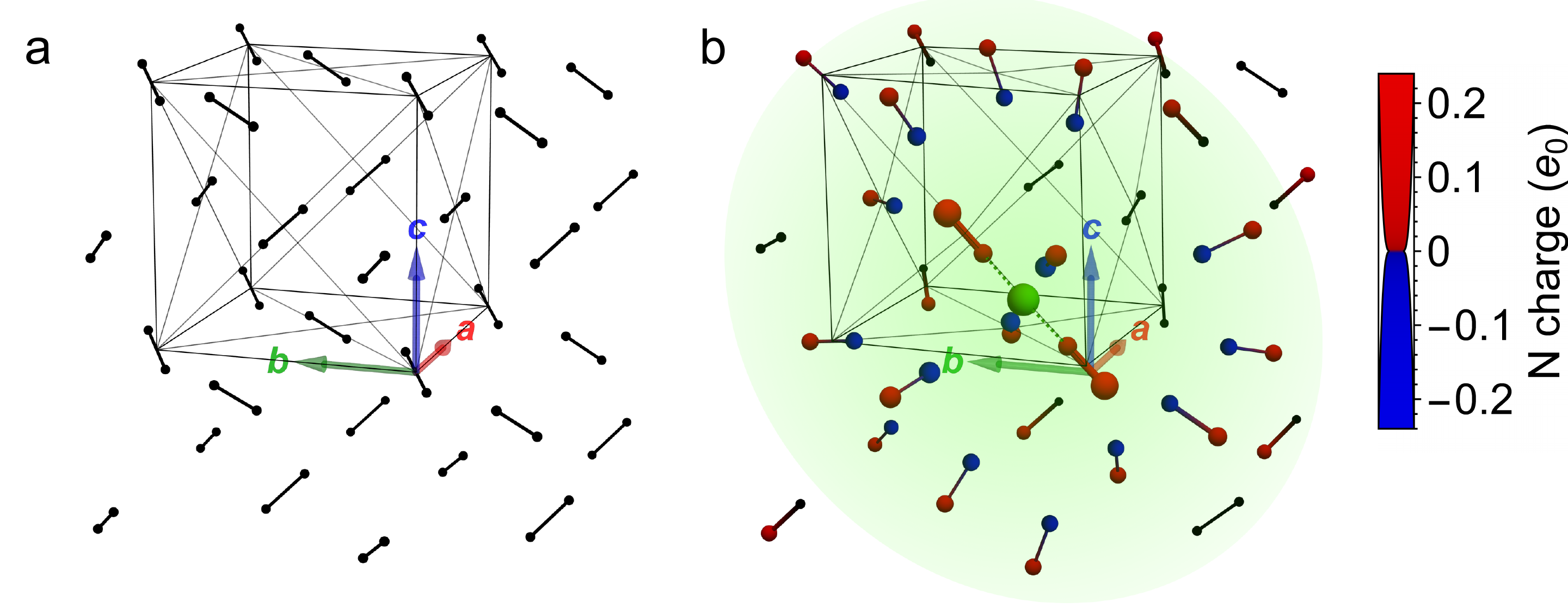}
\caption{%
{\bf Pristine and muonated crystal structures of solid nitrogen, \aNtwo.} 
Alternative views of {\bf a} the pristine \aNtwo structure, where the centers of \Ntwo molecules form a \fccLong lattice (gray lines), and 
{\bf b} the classical muon site, where \muonPlus (green) forms a linear \NtwomuNtwo complex with the two neighboring \Ntwo molecules at the center of an electric-dipole polaron (shading) [\cf Fig.~1a, b in the main text]. 
Induced \mulliken charge of nitrogen ions from \dftLongFirst calculations is represented by sphere size and color. 
}
\label{fig_site_alt}
\end{figure}

\begin{figure}[H]
\centering
\includegraphics[width=0.9\columnwidth]{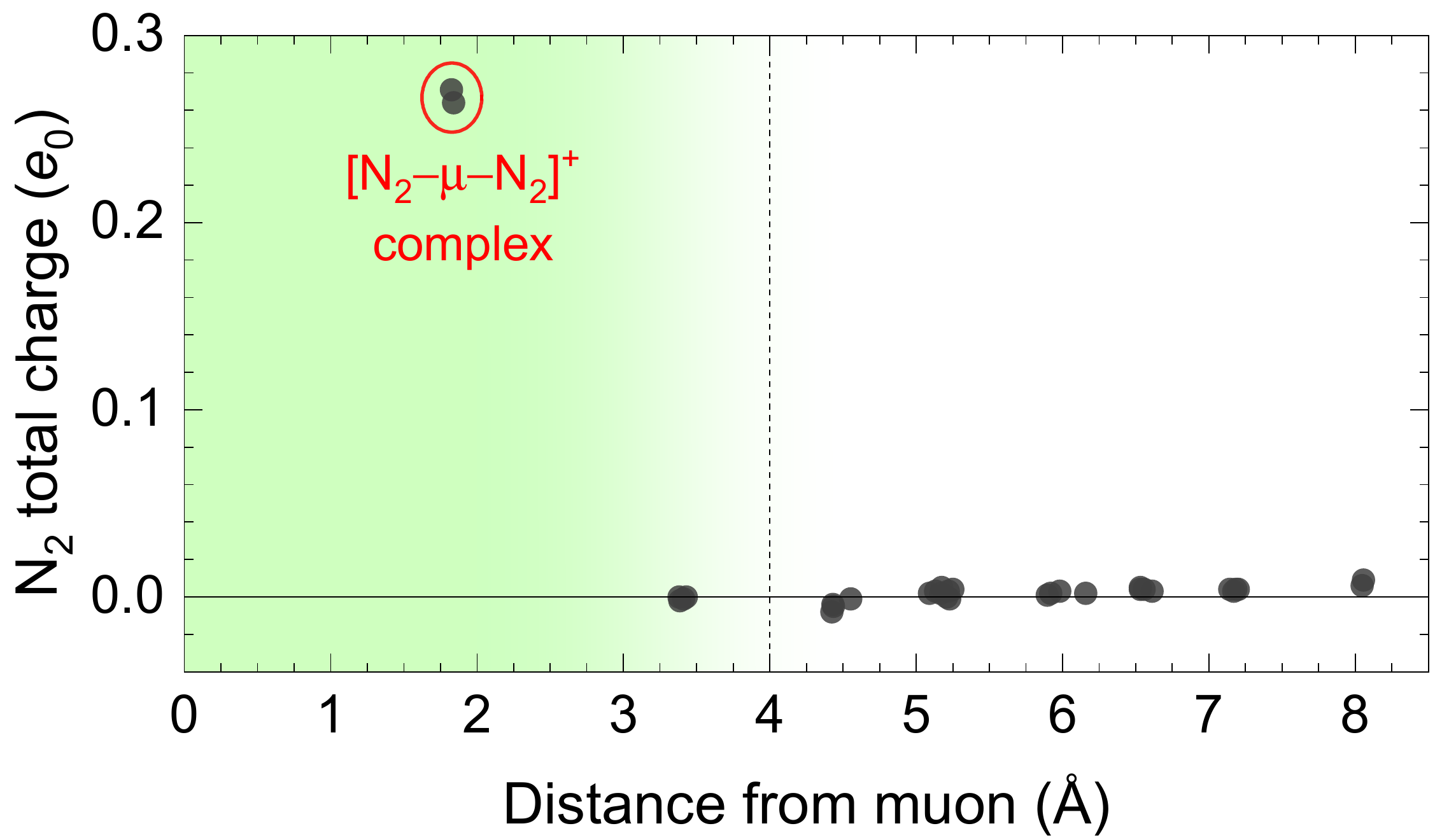}
\caption{%
{\bf Electric charge of \Ntwo molecules in muonated \aNtwo.} 
Net \mulliken electric charge on \Ntwo molecules in muonated \aNtwo for a classical muon site (see also Fig.~1b in the main text, \cref{fig_site_alt}b, and Supplementary Movie~2) from \dft. 
The only excess of net charge is seen within the \NtwomuNtwo complex due to polar covalent bonding of the two \Ntwo molecules with \muonPlus, which leaves them slightly positively charged while the muon's positive charge gets partially screened by negatively-charged electrons shared with the two \Ntwo. 
}
\label{fig_mol_charge}
\end{figure}

\begin{figure}[H]
\centering
\includegraphics[width=0.85\columnwidth]{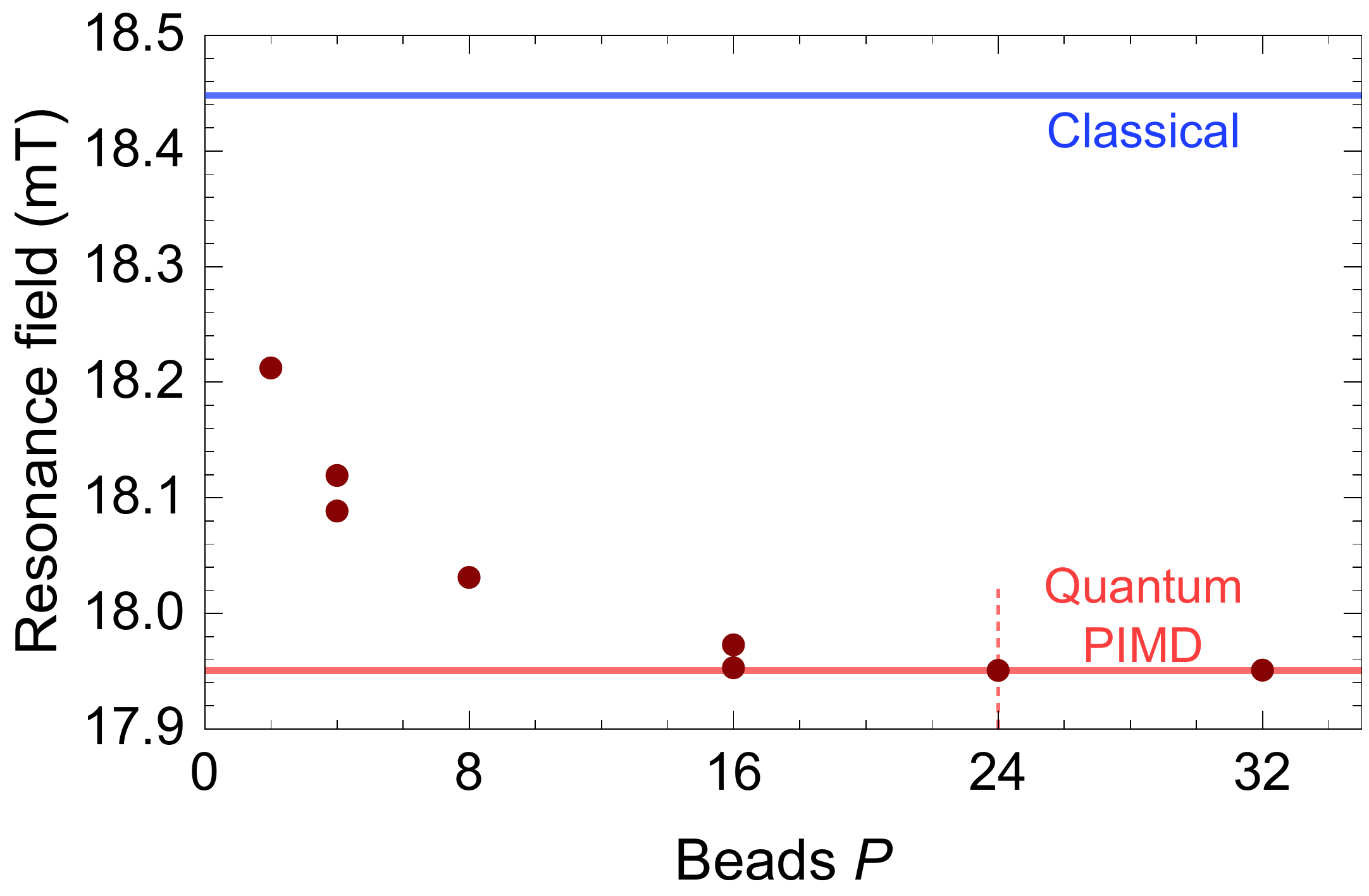}
\caption{%
{\bf \PimdLongFirst convergence.} 
Convergence of \pimd calculations with the number of beads $P$ as observed from the calibrated \qlcrLongFirst peak resonance field from a muon-lifetime weighed integral of the \qlcr signal over a $4$--$\SI{8}{\micro\second}$ time window. 
The final number of beads $P = 24$ for \pimd calculations presented in the main text is indicated by the dashed line. 
}
\label{fig_pimd_conv}
\end{figure}

\begin{figure}[H]
\centering
\includegraphics[width=0.9\columnwidth]{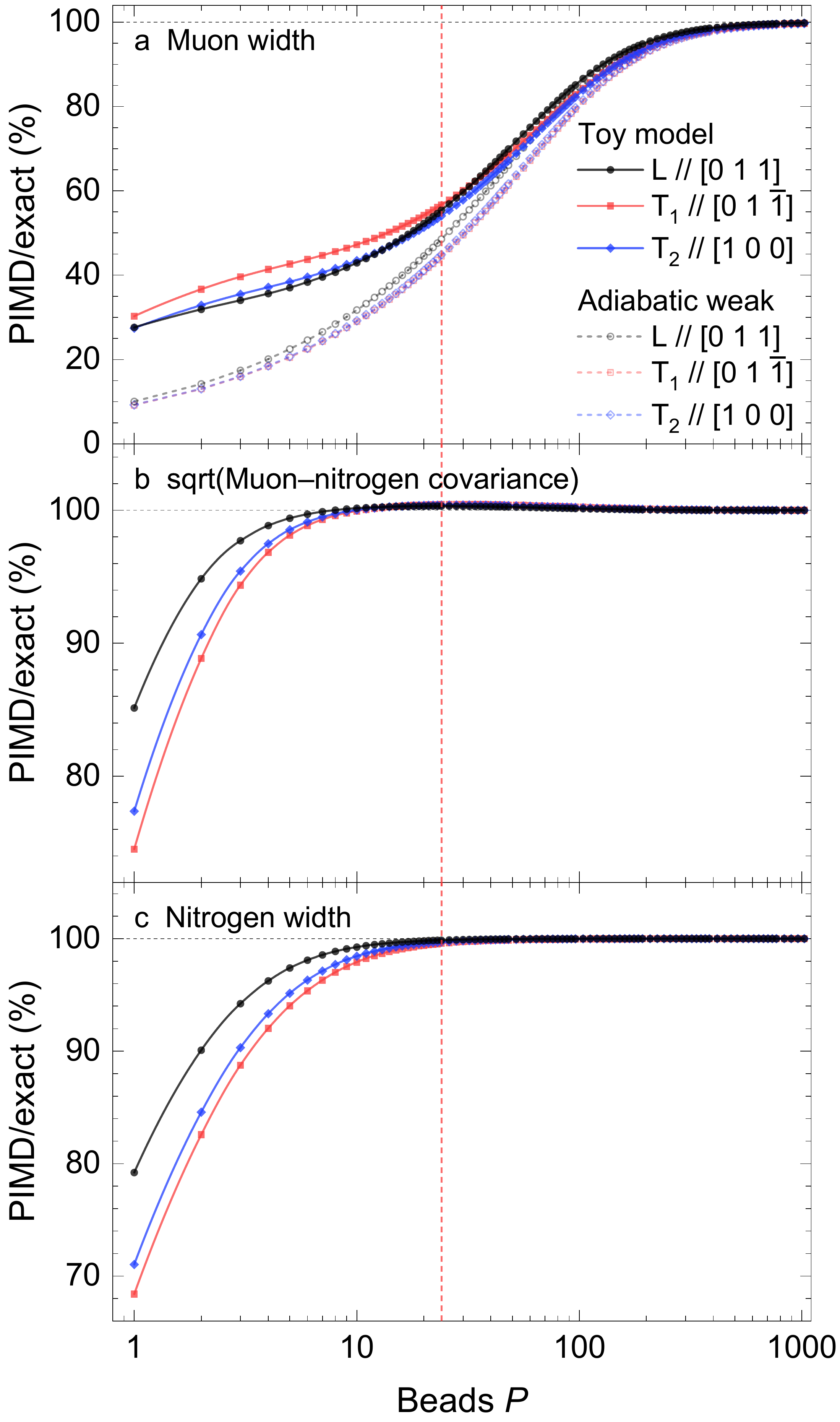}
\caption{%
{\bf Predicted \pimd convergence in the harmonic approximation.} 
Predicted convergence of \pimd calculations with the number of beads $P$ as observed from: {\bf a} the muon \wavefunction widths $\Delta x_j$, {\bf b} the covariances of muon and nitrogen positions, and {\bf c} the nitrogen \wavefunction widths along the $j = \mathrm{L}$, $\mathrm{T}_1$, and $\mathrm{T}_2$ directions. 
Solid lines are predictions of the toy model with $N = 2$ nearest nuclei described in the main text, while dashed lines represent predictions under the harmonic weakly-bound adiabatic approximation, \ie, for a single muon in an effective harmonic potential of Fig.~2a,b. 
The final number of beads $P = 24$ for \pimd calculations is indicated by the vertical dashed line. 
}
\label{fig_pimd_conv_widths}
\end{figure}


%
\begin{figure}[H]
\centering
\includegraphics[width=0.9\columnwidth]{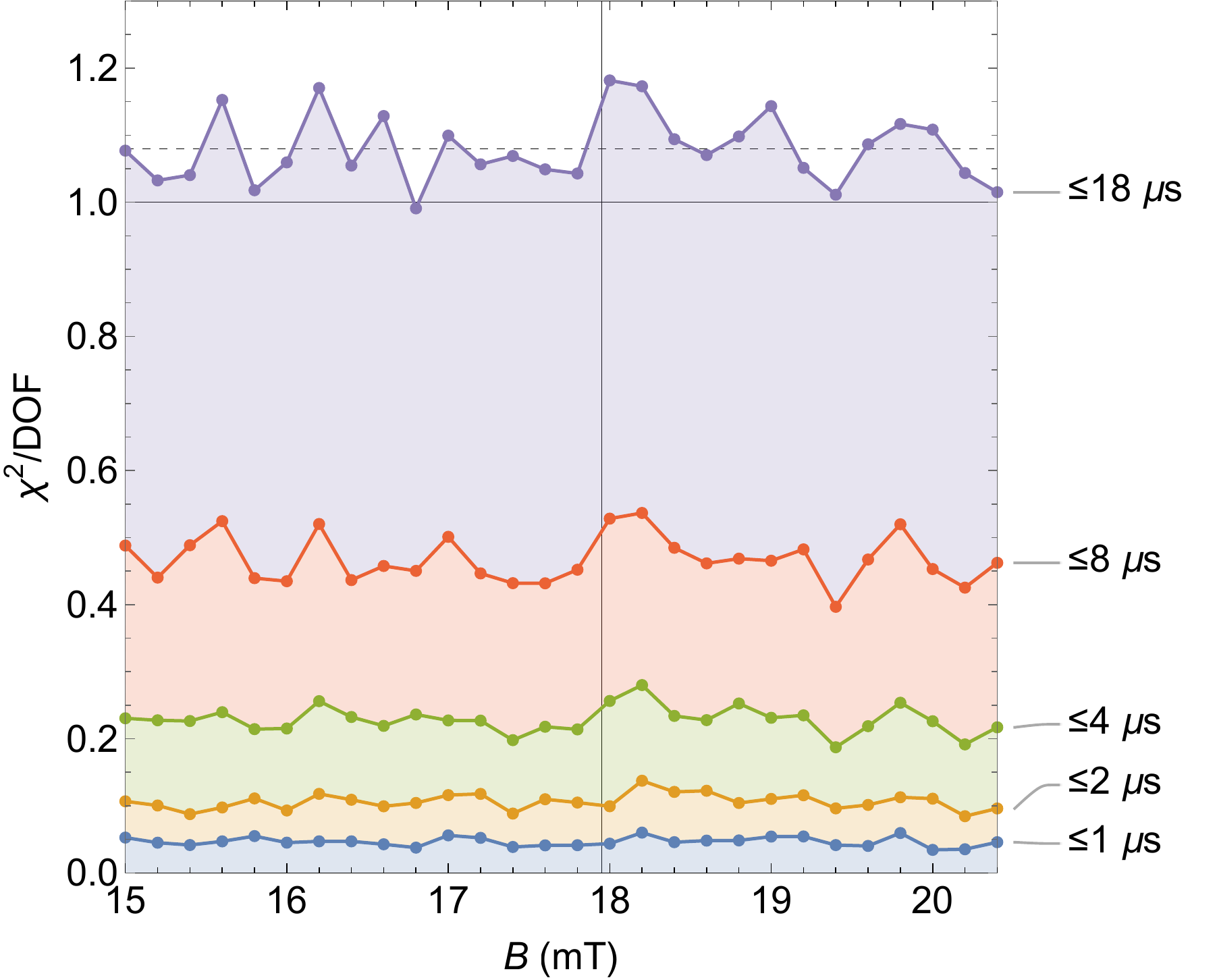} 
\caption{%
{\bf Quality of \qlcr spectra fits.} 
Field- and time-resolved quality of the global fits of experimental \qlcr spectra using \pimd at $T = \SI{5}{\kelvin}$ as described in the main text. 
The contributions towards the total reduced \reducedChiSq at each applied field $B$ are shown cumulatively up to different times after muon implantation. 
The horizontal dashed line shows the average $\reducedChiSqEq = 1.08$ across all $B$ (\ie, the total $\reducedChiSqEq$ of the global fit), while the vertical line shows the main \qlcr resonance field (\cref{fig_pimd_conv}). 
}
\label{fig_qlcr_fit_quality}
\end{figure}

\begin{figure}[H]
\centering
\includegraphics[width=1\columnwidth]{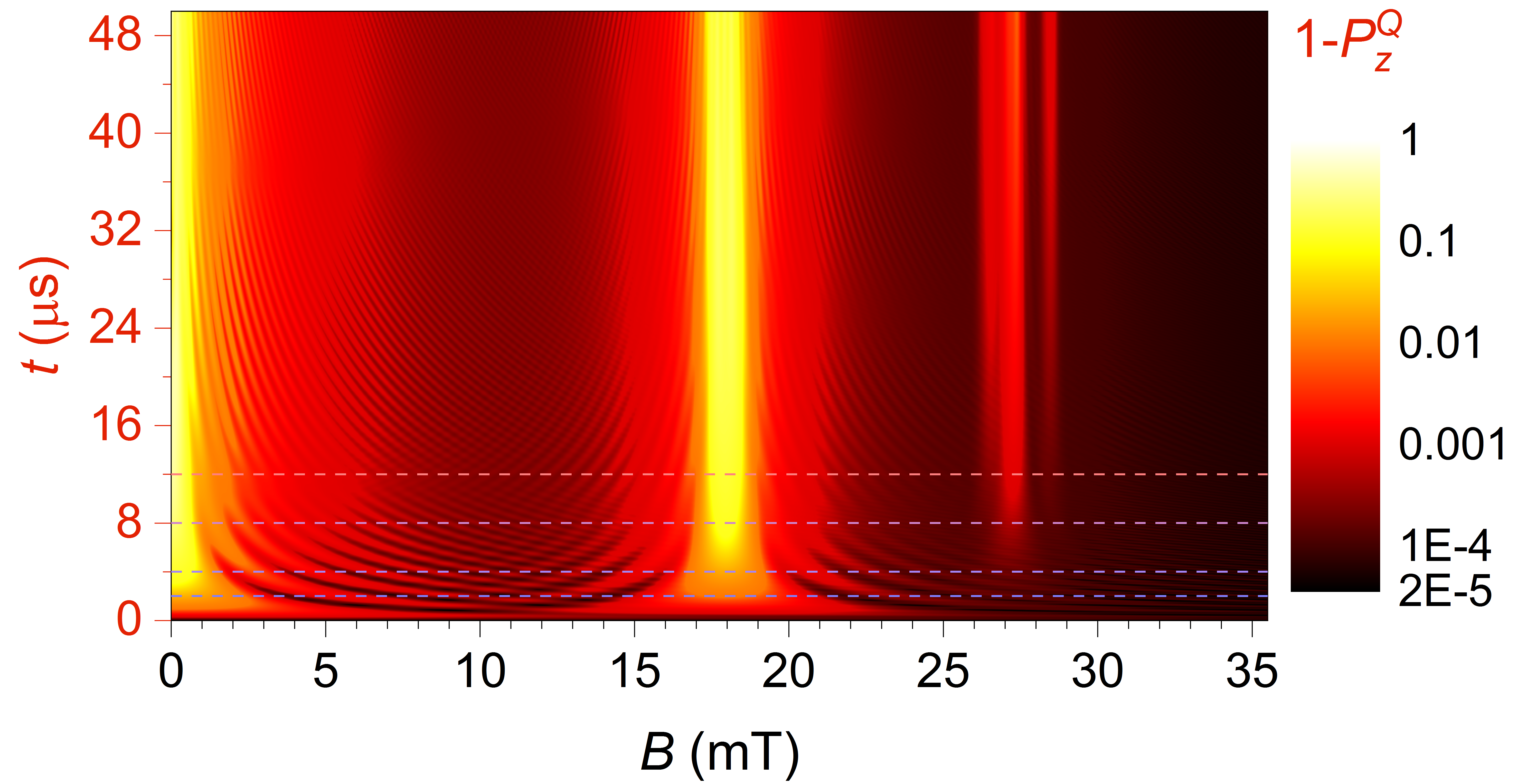}
\caption{%
{\bf Time-dependent \pimd \qlcr spectra.} 
Calibrated \qlcr spectra, $1 - P_\mathrm{z}^\mathrm{Q}$, from \pimd calculations with $24$ beads for time-differential \qlcr analysis. 
Dashed lines indicate time windows boundaries for \qlcr signal integrals shown in Fig.~3b in the main text. 
}
\label{fig_qlcr_signal}
\end{figure}

~
